\documentclass[11pt]{article}
\usepackage{amssymb}
\usepackage[dvips]{epsfig}
\usepackage{graphicx}
\makeatletter\@addtoreset {equation}{section}\makeatother
\newtheorem{theorem}{Theorem}[section]
\newtheorem{lemma}[theorem]{Lemma}
\newtheorem{corollary}[theorem]{Corollary}

\newtheorem{remark}[theorem]{Remark}
\newenvironment{proof}{
    \noindent {\it Proof.}}{\hfill$\Box$
}

\begin{document}

\title{\bf Two-pulse solutions in the fifth-order KdV equation :
rigorous theory and numerical approximations}
\author{Marina Chugunova and Dmitry Pelinovsky \\
Department of Mathematics, McMaster University, Hamilton, Ontario,
Canada, L8S 4K1}
\date{\today}
\maketitle

\begin{abstract}
We revisit existence and stability of two-pulse solutions in the
fifth-order Korteweg--de Vries (KdV) equation with two new
results. First, we modify the Petviashvili method of successive
iterations for numerical (spectral) approximations of pulses and
prove convergence of iterations in a neighborhood of two-pulse
solutions. Second, we prove structural stability of embedded
eigenvalues of negative Krein signature in a linearized KdV
equation. Combined with stability analysis in Pontryagin spaces,
this result completes the proof of spectral stability of the
corresponding two-pulse solutions. Eigenvalues of the linearized
problem are approximated numerically in exponentially weighted
spaces where embedded eigenvalues are isolated from the continuous
spectrum. Approximations of eigenvalues and full numerical
simulations of the fifth-order KdV equation confirm stability of
two-pulse solutions related to the minima of the effective
interaction potential and instability of two-pulse solutions
related to the maxima points.
\end{abstract}

\section{Introduction}

One-pulse solutions (solitons) are commonly met in many nonlinear
evolution equations where dispersive terms (represented by unbounded
differential operators) and nonlinear terms (represented by power
functions) are taken in a certain balance. Typical examples of such
nonlinear evolution equations with one-pulse solutions are given by
the NLS (nonlinear Schr\"{o}dinger) equation, the Klein-Gordon
(nonlinear wave) equation and the KdV (Korteweg-de Vries) equation,
as well as their countless generalizations.

One-pulse solutions are the only stationary (traveling) localized
solutions of the simplest nonlinear evolution equations. However,
uniqueness is not a generic property and bound states of spatially
separated pulses can represent other stationary (traveling)
localized solutions of the same evolution equation. For instance,
two-pulse, three-pulse, and generally $N$-pulse solutions exist in
nonlinear evolution equations with a higher-order dispersion
(represented by a higher-order differential operator). The
prototypical example of such situation is the fifth-order KdV
equation in the form,
\begin{equation}
\label{kdv} u_t + u_{xxx} - u_{xxxxx} + 2 u u_x = 0, \qquad x \in
\mathbb{R}, \quad t \in \mathbb{R}_+,
\end{equation}
where $u : \mathbb{R} \times \mathbb{R}_+ \mapsto \mathbb{R}$ and
all coefficients of the nonlinear PDE are normalized by a scaling
transformation. See Bridges \& Derks \cite{BD02} for a review of
history and applications of the fifth-order KdV equation (\ref{kdv})
to magneto-acoustic waves in plasma and capillary-gravity water
waves.

Traveling localized solutions $u(x,t) = \phi(x-ct)$ of the
fifth-order KdV equation (\ref{kdv}) satisfies the fourth-order
ODE
\begin{equation}
\label{ODE} \phi^{({\rm iv})} - \phi'' + c \phi = \phi^2, \qquad z
\in \mathbb{R},
\end{equation}
where $z = x-ct$ is the traveling coordinate and one integration of
the fifth-order ODE in $z$ is performed subject to zero boundary
conditions on $\phi(z)$ and its derivatives as $|z| \to \infty$.
Existence of localized solutions (homoclinic orbits) to the
fourth-order ODE (\ref{ODE}) was considered by methods of the
dynamical system theory. See Champneys \cite{C98} for a review of
various results on existence of homoclinic orbits in the ODE
(\ref{ODE}).

In particular, it is proved with the variational method by Buffoni
\& Sere \cite{BS96} and Groves \cite{G98} (see references to
earlier works in \cite{C98}) that the fourth-order ODE (\ref{ODE})
has the one-pulse solution $\phi(z)$ for $c > 0$, which is the
only localized solution of the ODE (\ref{ODE}) for $0 < c <
\frac{1}{4}$ up to the translation $\phi(z-s)$ for any $s \in
\mathbb{R}$. The analytical expression for the one-pulse solution
is only available for $c = \frac{36}{169} < \frac{1}{4}$ with
\begin{equation}
\label{nexact} \phi(z) = \frac{105}{338} \; {\rm
sech}^4\left(\frac{z}{2 \sqrt{13}}\right).
\end{equation}
For $c > \frac{1}{4}$, the fourth-order ODE (\ref{ODE}) has
infinitely many multi-pulse solutions in addition to the one-pulse
solution \cite{BS96,G98}. The multi-pulse solutions look like
multiple copies of the one-pulse solutions separated by finitely
many oscillations close to the zero equilibrium $\phi = 0$.
Stability and evolution of multi-pulse solutions are beyond the
framework of the fourth-order ODE (\ref{ODE}) and these questions
were considered by two equivalent theories in the recent past.

The pioneer work of Gorshkov \& Ostrovsky explains multi-pulse
solutions of the fifth-order KdV equation (\ref{kdv}) from the
effective interaction potential computed from the one-pulse
solution \cite{GO81,GOP84}. When the interaction potential has an
alternating sequence of maxima and minima (which corresponds to
the case when the one-pulse solution $\phi(z)$ has oscillatory
decaying tails at infinity), an infinite countable sequence of
two-pulse solutions emerge with the distance between the pulses
near the sequence of extremal points. Three-pulse solutions can be
constructed as a bi-infinite countable sequence of three one-pulse
solutions where each pair of two adjacent pulses is located
approximately at a distance defined by the two-pulse solution.
Similarly, $N$-pulse solutions can be formed by a $(N-1)$-infinite
countable sequence of $N$ copies of one-pulse solutions. The
perturbative procedure in \cite{GO81} has the advantages that both
the linear and nonlinear stability of multi-pulse solutions can be
predicted from analysis of the approximate ODE system derived for
distances between the individual pulses. Numerical evidences of
validity of this procedure in the context of the fifth-order KdV
equation are reported in \cite{CB97}.

A different theory was developed by Sandstede \cite{S98} who
extended the Lin's work on the Lyapunov--Schmidt reductions for
nonlinear evolution equations \cite{Lin90}. In this method, a
linear superposition of $N$ one-pulse solutions $\phi(z) =
\sum_{j=1}^N \Phi(z - s_j)$ is a solution of the ODE (\ref{ODE})
in the case when the distances between pulses are infinite (i.e.
$|s_{j+1} - s_j| = \infty$, $\forall j$). The Jacobian of the
nonlinear ODE (\ref{ODE}) defines a linear self-adjoint operator
from $H^2(\mathbb{R})$ to $L^2(\mathbb{R})$:
\begin{equation}
\label{Jacobian} {\cal H} = c - \partial^2_z + \partial^4_z - 2
\phi(z), \qquad c > 0,
\end{equation}
where the unbounded differential part $c - \partial^2_z +
\partial^4_z$ is positive and bounded away from zero while the
exponentially decaying potential term $-2 \phi(z)$ is a relatively
compact perturbation. When $\phi(z)$ is a linear superposition of
$N$ infinitely-separated one-pulse solutions $\Phi(z-s_j)$, the
Jacobian ${\cal H}$ has $N$ zero eigenvalues related to the
eigenfunctions $\Phi'(z-s_j)$ due to the translational invariance
of the ODE (\ref{ODE}). The Lyapunov--Schmidt method leads to a
system of bifurcation equations for the distances between
individual pulses. When $\phi(z)$ is the $N$-pulse solution with
finitely separated pulses (i.e. $|s_{j+1} - s_j| < \infty$,
$\forall j$), one zero eigenvalue of the Jacobian operator ${\cal
H}$ survives beyond the reductive procedure due to the
translational invariance of the $N$-pulse solution $\phi(z)$,
while $N-1$ real eigenvalues bifurcate from zero. The reduction
method gives not only information about existence of multi-pulse
solutions but also prediction of their spectral stability in the
linearized time-evolution problem \cite{S98}. The linearized
problem for the fifth-order KdV equation takes the form
\begin{equation}
\label{stability-problem} \partial_z {\cal H} v = \lambda v,
\qquad z \in \mathbb{R},
\end{equation}
where $v : \mathbb{R} \mapsto \mathbb{C}$ is an eigenfunction for
a small perturbation of $\phi(z)$ in the reference frame $z = x -
ct$ and $\lambda \in \mathbb{C}$ is an eigenvalue. We say that the
eigenvalue $\lambda$ is {\em unstable} if ${\rm Re}(\lambda) > 0$
and $v \in H^2(\mathbb{R})$. We say that the eigenvalue $\lambda$
is of {\em negative Krein signature} if ${\rm Re}(\lambda) = 0$,
${\rm Im}(\lambda) > 0$, $v \in H^2(\mathbb{R})$ and $({\cal H} v,
v) < 0$.

Our interest to this well-studied problem is revived by the recent
progress in the spectral theory of non-self-adjoint operators
arising from linearizations of nonlinear evolution equations
\cite{ChPel}. These operators can be defined as self-adjoint
operators into Pontryagin space where they have a
finite-dimensional negative invariant subspace. Two physically
relevant problems for the fifth-order KdV equation (\ref{kdv})
have been solved recently by using the formalism of operators in
Pontryagin spaces. First, convergence of the numerical iteration
method (known as the Petviashvili's method) for one-pulse
solutions of the ODE (\ref{ODE}) was proved by the contraction
mapping principle in a weighted Hilbert space (which is equivalent
to Pontryagin space with zero index) \cite{PelStep}. Second,
eigenvalues of the spectral stability problem in a linearization
of the fifth-order KdV equation (\ref{kdv}) were characterized in
Pontryagin space with a non-zero index defined by the finite
number of negative eigenvalues of ${\cal H}$ \cite{KodPel,ChPel}.

Both recent works rise some open problems when the methods are
applied to the $N$-pulse solutions in the fifth-order KdV equation
(\ref{kdv}), even in the case of two-pulse solutions ($N=2$). The
successive iterations of the Petviashvili's method do not converge
for two-pulse solutions. The iterative sequence with two pulses
leads either to a single pulse or to a spurious solution with two
pulses located at an arbitrary distance (see Remark 6.5 in
\cite{PelStep}). This numerical problem arises due to the presence
of small and negative eigenvalues of ${\cal H}$. A modification of
the Petviashvili's method is needed to suppress these eigenvalues
similarly to the work of Demanet \& Schlag \cite{DemSchlag} where
the zero eigenvalue associated to the translational invariance of
the three-dimensional NLS equation is suppressed. We shall present
the modification of the iterative Petviashvili's method in this
article.

Our numerical method complements the recent work of Yang \& Lakoba
\cite{YangLakoba} where the steepest descent method is developed
to suppress any unstable eigenvalues of the iterative method and
to enforce its convergence. We note that other techniques for
numerical approximation of multi-pulse solutions in the
fifth-order KdV equation have been developed by Champneys and
coworkers \cite{Champneys0,Champneys} on the basis of a delicate
combination of the numerical shooting method and the continuation
techniques.

Open questions also arise when spectral stability of multi-pulse
solutions is considered within the linear eigenvalue problem
(\ref{stability-problem}). By either the Gorshkov--Ostrovsky
perturbative procedure or the Sandstede--Lin reduction method, the
small eigenvalues of the Jacobian operator ${\cal H}$ result in
small eigenvalues of the linearized operator $\partial_z {\cal
H}$, which are either pairs of real eigenvalues (one of which is
unstable) or pairs of purely imaginary eigenvalues of negative
Krein signature (which are neutrally stable but potentially
unstable). Both cases are possible in the fifth-order KdV equation
in agreement with the count of unstable eigenvalues in Pontryagin
spaces (see Theorem 6 in \cite{ChPel}). (Similar count of unstable
eigenvalues and eigenvalues of negative Krein signatures was
developed for the NLS equations in recent papers
\cite{KKS04,Pel05}.) Since the real eigenvalues are isolated from
the continuous spectrum of the eigenvalue problem
(\ref{stability-problem}), they are structurally stable and
persist with respect to parameter continuations. However, the
purely imaginary eigenvalues are embedded into the continuous
spectrum of the eigenvalue problem (\ref{stability-problem}) and
their destiny remains unclear within the reduction methods. It is
well known for the NLS-type and Klein--Gordon-type equations that
embedded eigenvalues are structurally unstable to the parameter
continuations \cite{G90}. If a certain Fermi golden rule related
to the perturbation term is nonzero, the embedded eigenvalues of
negative Krein signature bifurcate off the imaginary axis to
complex eigenvalues inducing instabilities of pulse solutions
\cite{CPV05}. (The embedded eigenvalues of positive Krein
signature simply disappear upon a generic perturbation
\cite{CPV05}.) This bifurcation does not contradict the count of
unstable eigenvalues \cite{KKS04,Pel05} and it is indeed observed
in numerical approximations of various pulse solutions of the
coupled NLS equations \cite{PelYang05}.

From a physical point of view, we would expect that the time
evolution of an energetically stable superposition of stable
one-pulse solutions remains stable. (Stability of one-pulse
solutions in the fifth-order KdV equation (\ref{kdv}) was
established with the variational theory \cite{L99} and the
multi-symplectic Evans function method \cite{BD02,BDG02}.)
According to the Gorshkov-Ostrovsky perturbative procedure,
dynamics of well-separated pulses is represented by the Newton
particle's law which describes nonlinear stability of oscillations
near the minima of the effective interaction potential
\cite{GOP84}. Therefore, we would rather expect (on the contrary
to embedded eigenvalues in the linearized NLS and Klein--Gordon
equations) that the embedded eigenvalues of negative Krein
signature are structurally stable in the linear eigenvalue problem
(\ref{stability-problem}) and persist beyond the leading order of
the perturbative procedure. (Multi-pulse solutions of the NLS and
Klein--Gordon equations with well-separated individual pulses are
always linearly stable since the small purely imaginary
eigenvalues of the Lyapunov--Schmidt reductions are isolated from
the continuous spectrum of the corresponding linearized problems
\cite{Yew}.)

Since the count of unstable eigenvalues in \cite{ChPel} does not
allow us to prove structural stability of embedded eigenvalues of
negative Krein signature, we address this problem separately by
using different analytical and numerical techniques. In
particular, we present an analytical proof of persistence
(structural stability) of embedded eigenvalues of negative Krein
signature in the linearized problem (\ref{stability-problem}). We
also apply the Fourier spectral method and illustrate the
linearized stability of the corresponding two-pulse solutions
numerically. Our analytical and numerical methods are based on the
construction of exponentially weighted spaces for the linear
eigenvalue problem (\ref{stability-problem}). (See \cite{PW94} for
analysis of exponentially weighted spaces in the context of the
generalized KdV equation.) Our results complement the recent work
of Sandstede  \cite{Sandstede} and Bridges et al. \cite{Dias},
where two-pulse solutions of the fifth-order KdV equation
(\ref{kdv}) were also addressed. Additionally, Lyapunov--Schmidt
reductions of two-pulse solutions of the coupled KdV equations
were recently considered by Scheel \& Wright \cite{SchWright}.

This article is structured as follows. Section 2 contains summary
of available results on existence and stability of one-pulse and
two-pulse solutions of the fifth-order KdV equation (\ref{kdv}).
Section 3 presents a modification of the iterative Petviashvili
method for convergent numerical approximations of the two-pulse
solutions in the fourth-order ODE (\ref{ODE}). Section 4 develops
the proof of structural stability of embedded eigenvalues in the
eigenvalue problem (\ref{stability-problem}) and numerical
approximations of unstable and stable eigenvalues in an
exponentially weighted space. Section 5 describes full numerical
simulations of the fifth-order KdV equation (\ref{kdv}) to study
nonlinear dynamics of two-pulse solutions.

\section{Review of available results on two-pulse solutions}

Linearization of the ODE (\ref{ODE}) at the critical point
$(0,0,0,0)$ leads to the eigenvalues $\kappa$ given by roots of the
quartic equation,
\begin{equation}
\label{quartic} \kappa^4 - \kappa^2 + c = 0.
\end{equation}
When $c < 0$, one pair of roots $\kappa$ is purely imaginary and
the other pair is purely real. When $0 < c < \frac{1}{4}$, two
pairs of roots $\kappa$ are real-valued. When $c > \frac{1}{4}$,
the four complex-valued roots $\kappa$ are located symmetric about
the axes. We will use notations $k_0 = {\rm Im}(\kappa) > 0$ and
$\kappa_0 = {\rm Re}(\kappa) > 0$ for a complex root of
(\ref{quartic}) in the first quadrant for $c > \frac{1}{4}$. The
following two theorems summarize known results on existence of
one-pulse and two-pulse solutions of the ODE (\ref{ODE}).

\begin{theorem}[One-pulse solutions]
Consider the ODE (\ref{ODE}) with $c > 0$.
\begin{itemize}
\item[(i)] There exists an isolated one-pulse solution $\phi(z)$
such that $\phi \in H^2(\mathbb{R}) \cap C^5(\mathbb{R})$, $\phi(-z)
= \phi(z)$, and $\phi(z) \to 0$ exponentially as $|z| \to \infty$.
Moreover, $\phi(z)$ is $C^m(\mathbb{R})$ for any $m \geq 0$.

\item[(ii)] The Jacobian operator ${\cal H}$ associated with the one-pulse
solution $\phi(z)$ has exactly one negative eigenvalue with an even
eigenfunction and a simple kernel with the odd eigenfunction
$\phi'(z)$.

\item[(iii)] Assume that the map $\phi(z)$ from $c > 0$ to
$H^2(\mathbb{R})$ is $C^1(\mathbb{R}_+)$. Then $P'(c) > 0$, where
$P(c) = \| \phi \|_{L^2}^2$. The linearized operator $\partial_z
{\cal H}$ has a two-dimensional algebraic kernel in
$L^2(\mathbb{R})$ and no unstable eigenvalues with ${\rm
Re}(\lambda)
> 0$.
\end{itemize}
\label{theorem-one-pulse}
\end{theorem}

\begin{proof}
(i) Existence of a symmetric solution $\phi(z)$ in
$H^2(\mathbb{R})$ follows by the mountain-pass lemma and
concentration--compactness principle (see Theorem 8 in \cite{G98}
and Theorem 2.3 in \cite{L99}). The equivalence between weak
solutions of the variational theory and strong solutions of the
ODE (\ref{ODE}) is established in Lemma 1 of \cite{G98} and Lemma
2.4 of \cite{L99}. The exponential decay of $\phi(z)$ follows by
the Stable Manifold Theorem in Appendix A of \cite{BS96}. Global
uniqueness of the symmetric solution $\phi(z)$ for $0 < c <
\frac{1}{4}$ (module the space translation in $z$) is proved in
\cite{AT91}. Local uniqueness of the symmetric one-pulse solution
$\phi(z)$ for $c \geq \frac{1}{4}$ follows from the fact that the
Jacobian operator ${\cal H}$ has a simple kernel due to the
translational invariance (statement (ii)). Finally, the smoothness
of the function $\phi(z)$ is proved from the ODE (\ref{ODE}) by
the bootstrapping principle \cite{CCP06}.

(ii) The Jacobian operator ${\cal H}$ coincides with the Hessian
of the energy functional under a fixed value of the momentum
functional \cite{BSS87}. Uniqueness of the negative eigenvalue of
${\cal H}$ follows from the fact that the one-pulse solution
$\phi(z)$ is a ground state (a global minimizer) of the
constrained manifold (see Proposition 16 in \cite{G98}). The
ground state for operators with even derivatives and symmetric
potential functions is always given by the even eigenfunction. The
kernel of ${\cal H}$ is simple due to the duality principle in
Theorem 4.1 of \cite{BS96}. If it is not simple, then global
two-dimensional stable and unstable manifolds coincide and the
time for a homoclinic orbit to go from the local unstable manifold
to the local stable manifold is uniformly bounded. However, it was
shown in \cite{B96} that a sequence of homoclinic solutions $\{
u_n \}_{n \in \mathbb{N}}$ exists such that the time between local
manifolds grows linearly in $n$. Hence, the kernel of ${\cal H}$
is simple with the odd eigenfunction $\phi'(z)$ due to the space
translation.

(iii) Smoothness of the map $\phi(z)$ from $c > 0$ to
$H^2(\mathbb{R})$ is a standard assumption (see Assumption 5.1 in
\cite{L99}). The positive value of $P'(c)$ and the absence of
eigenvalues of $\partial_z {\cal H}$ with ${\rm Re}(\lambda) > 0$
follow from the stability of one-pulse solutions in Main Theorem
of \cite{BSS87}, Theorems 4.1 and 5.3 of \cite{L99} and Theorem
8.1 of \cite{BD02}. Indeed, the only negative eigenvalue of ${\cal
H}$ becomes a positive eigenvalue in the constrained subspace
where the corresponding eigenfunction is orthogonal to $\phi(z)$
\cite{SS90}. The two-dimensional algebraic kernel of $\partial_z
{\cal H}$ follows from the derivatives of the ODE (\ref{ODE}) in
$z$ and $c$:
\begin{equation}
\label{algebraic-kernel} {\cal H} \phi'(z) = 0, \qquad {\cal H}
\partial_c \phi(z) = - \phi(z).
\end{equation}
The algebraic kernel of $\partial_z {\cal H}$ is exactly
two-dimensional under the condition $P'(c) \neq 0$ \cite{PW92}.
\end{proof}

\begin{theorem}[Two-pulse solutions]
Consider the ODE (\ref{ODE}) with $c > \frac{1}{4}$. There exists
an infinite countable set of two-pulse solutions $\phi(z)$ such
that $\phi \in H^2(\mathbb{R}) \cap C^5(\mathbb{R})$, $\phi(-z) =
\phi(z)$, $\phi(z) \to 0$ exponentially as $|x| \to \infty$, and
$\phi(z)$ resembles two copies of the one-pulse solutions
described in Theorem \ref{theorem-one-pulse} which are separated
by small-amplitude oscillatory tails. The members of the set are
distinguished by the distance $L$ between individual pulses which
takes the discrete values $\{ L_n\}_{n \in \mathbb{N}}$. Moreover,
for any small $\delta > 0$ there exists $\gamma > 0$ such that
\begin{equation}
\label{distance-distribution} \left| L_n - \frac{2 \pi n}{k_0} -
\gamma \right| < \delta, \qquad n \in \mathbb{N}.
\end{equation}
\label{theorem-two-pulses}
\end{theorem}

\begin{proof}
Existence of an infinite sequence of geometrically distinct
two-pulse solutions with the distances distributed by
(\ref{distance-distribution}) follows by the variational theory in
Theorem 1.1 of \cite{BS96} under the assumption that the
single-pulse solution $\phi(z)$ is isolated (up to the space
translations). This assumption is satisfied by Theorem
\ref{theorem-one-pulse}(ii).
\end{proof}

The following theorem describes an asymptotic construction of the
two-pulse solutions, which is used in the rest of our paper.

\begin{theorem}[Lyapunov--Schmidt reductions for two-pulse solutions] Let
$c > \frac{1}{4}$ and let $\Phi(z)$ denote the one-pulse solution
described by Theorem \ref{theorem-one-pulse}. Let $L = 2s$ be the
distance between two copies of the one-pulse solutions of the ODE
(\ref{ODE}) in the decomposition
\begin{equation}
\label{decomposition} \phi(z) = \Phi(z - s) + \Phi(z + s) +
\varphi(z),
\end{equation}
where $\varphi(x)$ is a remainder term. Let $W(L)$ be
$C^m(\mathbb{R}_+)$ function for any $m \geq 0$ defined by
\begin{equation}
\label{effective-potential} W(L) = \int_{\mathbb{R}} \Phi^2(z)
\Phi(z + L) dz.
\end{equation}
There exists an infinite countable set of extrema of $W(L)$, which
is denoted by $\{ L_n \}_{n \in \mathbb{N}}$.

\begin{itemize}
\item[(i)] Assume that $W''(L_n) \neq 0$ for a given $n \in
\mathbb{N}$. There exists a unique symmetric two-pulse solution
$\phi(z) \equiv \phi_n(z)$ described by Theorem
\ref{theorem-two-pulses}, such that the distance $L$ between
individual pulses satisfies the bound
\begin{equation}
|L - L_n| \leq C_n e^{-\kappa_0 L}
\end{equation}
for some $C_n > 0$.

\item[(ii)] The Jacobian ${\cal H}$ associated with the $n$-th two-pulse solution
$\phi_n(z)$ has exactly two finite negative eigenvalues with even
and odd eigenfunctions, a simple kernel with the odd eigenfunction
$\phi_n'(z)$ and a small eigenvalue $\mu$ with an even
eigenfunction, such that
\begin{equation}
\label{eigenvalues-H-asymptotic} \left| \mu + \frac{2
W''(L_n)}{Q(c)} \right| \leq \tilde{C}_n e^{- 2\kappa_0 L}
\end{equation}
for some $\tilde{C}_n > 0$, where $Q(c) = \| \Phi' \|^2_{L^2} > 0$.
In particular, the small eigenvalue $\mu$ is negative when $W''(L_n)
> 0$ and positive when $W''(L_n) < 0$.

\item[(iii)] There exists a pair of small eigenvalues $\lambda$ of
the linearized operator $\partial_z {\cal H}$ associated with the
$n$-th two-pulse solution $\phi_n(z)$, such that
\begin{equation}
\label{eigenvalues-asymptotic} \left| \lambda^2 + \frac{4
W''(L_n)}{P'(c)} \right| \leq \hat{C}_n e^{- 2 \kappa_0 L},
\end{equation}
for some $\hat{C}_n > 0$, where $P(c) = \| \Phi \|^2_{L^2}$ and
$P'(c) > 0$. In particular, the pair is real when $W''(L_n) < 0$
and purely imaginary (up to the leading order) with negative Krein
signature when $W''(L_n) > 0$.
\end{itemize}
\label{theorem-slow-dynamics}
\end{theorem}

\begin{proof}
When the tails of the one-pulse solution $\Phi(z)$ are decaying
and oscillatory (i.e. when $c > \frac{1}{4}$), the smooth function
$W(L)$ in (\ref{effective-potential}) is decaying and oscillatory
in $L$ and an infinite set of extrema $\{ L_n \}_{n \in
\mathbb{N}}$ exists. Let us pick $L_n$ for a fixed value of $n \in
\mathbb{N}$ such that $W'(L_n) = 0$ and $W''(L_n) \neq 0$.

(i) When the decomposition (\ref{decomposition}) is substituted
into the ODE (\ref{ODE}), one can find the ODE for $\varphi(z)$:
\begin{equation}
\label{PDE-kdv} \left( c - \partial^2_z + \partial^4_z - 2 \Phi(z -
s) - 2 \Phi(z+s) \right) \varphi - \varphi^2 = 2 \Phi(z - s)
\Phi(z+s).
\end{equation}
Let $\epsilon = {\rm O}(e^{-\kappa_0 L})$ be a small parameter
that measures the $L^{\infty}$-norm of the overlapping term
$\Phi(z-s) \Phi(z+s)$ in the sense that for each $\epsilon > 0$
there exist constants $C_0 > 0$ and $s_0 > 0$ such that
\begin{equation}
\label{small-parameter} \|\Phi(z - s) \Phi(z+s) \|_{L^{\infty}} \leq
C_0 \epsilon \qquad \forall s \geq s_0.
\end{equation}
Denote $L = 2s$ and $\epsilon \Psi(z,L) = 2 \Phi(z) \Phi(z+L)$ and
rewrite the ODE (\ref{PDE-kdv}) for $\tilde{\varphi}(z) =
\varphi(z+s)$:
\begin{equation}
\label{PDE-kdv-2} \left( c - \partial^2_z + \partial^4_z - 2 \Phi(z)
\right) \tilde{\varphi} - 2 \Phi(z+L)\tilde{\varphi}-
\tilde{\varphi}^2 = \epsilon \Psi(z,L).
\end{equation}
The ODE (\ref{PDE-kdv-2}) is defined in function space
$H^2(\mathbb{R})$, where the Jacobian for the one-pulse solution
$$
{\cal H} = c - \partial^2_z + \partial^4_z - 2 \Phi(z)
$$
has a simple kernel with the odd eigenfunction $\Phi'(z)$ by
Theorem \ref{theorem-one-pulse}(ii). By the Lyapunov--Schmidt
reduction method (see \cite{GS81}), there exists a unique solution
$\tilde{\varphi} = \tilde{\varphi}(z,L,\epsilon) \in
H^2(\mathbb{R}) \backslash {\rm ker}({\cal H})$, such that
$\tilde{\varphi}(z,L,0) = 0$ and $\tilde{\varphi}(z,L,\epsilon)$
is smooth in $\epsilon$, provided $L$ solves the bifurcation
equation $F(L,\epsilon) = 0$, where
\begin{eqnarray}
\nonumber F(L,\epsilon) & = & \epsilon \left( \Phi'(z),\Psi(z,L)
\right) + 2 \left( \Phi'(z),\Phi(z+L) \tilde{\varphi}(z) \right) +
\left( \Phi'(z), \tilde{\varphi}^2(z) \right) \\
\label{LS-bifurcation-equation} & = & -\left( \Phi^2(z),\partial_z
\Phi(z+L) \right) - \epsilon \left( \partial_L \Psi(z,L),
\tilde{\varphi}(z) \right) - \epsilon \left(\Psi(z,L),
\partial_z \tilde{\varphi}(z) \right) + \left( \Phi'(z),
\tilde{\varphi}^2(z) \right).
\end{eqnarray}
Since $\tilde{\varphi}(z,L,\epsilon)$ is smooth in $\epsilon$ with
$\tilde{\varphi}(z,L,0) = 0$, we obtain the expansion
$$
F(L,\epsilon) = - W'(L) + \tilde{F}(L,\epsilon),
$$
where the first term has the order of ${\rm O}(\epsilon)$ and the
second term has the order of ${\rm O}(\epsilon^2)$. The statement
follows by the Implicit Function Theorem applied to the scalar
equation $F(L,\epsilon) = 0$ under the assumption that the root
$L_n$ of $W'(L)$ is simple.

(ii) The Jacobian ${\cal H}$ associated with the $n$-th two-pulse
solution $\phi(z) \equiv \phi_n(z)$ given by the decomposition
(\ref{decomposition}) has the form:
$$
{\cal H} = c - \partial_z^2 + \partial_z^4 - 2 \Phi(z-s) - 2
\Phi(z+s) - 2 \varphi(z)
$$
In the limit of $\epsilon = 0$, where $\epsilon = {\rm
O}(e^{-\kappa_0 L})$, the Jacobian ${\cal H}$ has a double
negative eigenvalue and a double zero eigenvalue. By a linear
combination of eigenfunctions, one can construct one even and one
odd eigenfunctions for each of the double eigenvalues. By
continuity of eigenvalues of self-adjoint operators, the double
negative eigenvalue splits but the two negative eigenvalues remain
negative if $\epsilon$ is small. By reversibility of the system,
eigenfunctions for simple eigenvalues are either even or odd and
by continuity of eigenfunctions, there is exactly one even and one
odd eigenfunctions for the two negative eigenvalues. By the
translation invariance, the double zero eigenvalue splits into a
simple zero eigenvalue which corresponds to the odd eigenfunction
$\phi_n'(z)$ and to a small non-zero eigenvalue of the order of
${\rm O}(\epsilon)$ which corresponds to an even eigenfunction.
The splitting of the double zero eigenvalue in the problem ${\cal
H} v = \mu v$ is considered by the perturbation theory,
\begin{equation}
\label{eigenvector-0} v(z) = \alpha_1 \Phi'(z-s) + \alpha_2
\Phi'(z+s) + V(z),
\end{equation}
where $(\alpha_1,\alpha_2)$ are coordinates of the projections to
the kernel of ${\cal H}$ at $\epsilon = 0$ and $V(z)$ is the
remainder term of the order of ${\rm O}(\epsilon)$. By projecting
the eigenvalue problem ${\cal H} v = \mu v$ to the kernel of
${\cal H}$ and neglecting the terms of the order of ${\rm
O}(\epsilon^2)$, one can obtain a reduced eigenvalue problem:
\begin{eqnarray*}
\mu Q(c) \alpha_1 = -\tilde{W} \alpha_1 + W''(L_n) \alpha_2, \qquad
\mu Q(c) \alpha_2 = W''(L_n) \alpha_1 -\tilde{W} \alpha_2,
\end{eqnarray*}
where $Q(c) = \| \Phi' \|^2_{L^2} > 0$, $W''(L_n)$ is computed from
(\ref{effective-potential}) and
$$
\tilde{W} = 2 \left( [\Phi'(z - s)]^2, \varphi(z) + \Phi(z + s)
\right) = 2 \left( [\Phi'(z)]^2, \tilde{\varphi}(z) + \Phi(z + L)
\right).
$$
Since one eigenvalue must be zero with the odd eigenfunction
$\phi_n'(z)$, the zero eigenvalue corresponds to the eigenfunction
(\ref{eigenvector-0}) with $\alpha_1 = \alpha_2$ up to the leading
order. By looking at the linear system, we find that the zero
eigenvalue corresponding to $\alpha_1 = \alpha_2$ exists only if
$\tilde{W} = W''(L_n)$. The other eigenvalue at the leading order
is $\mu = - 2W''(L_n)/Q(c)$. The non-zero eigenvalue of the
reduced eigenvalue problem corresponds to the even eigenfunction
(\ref{eigenvector-0}) with $\alpha_1 = - \alpha_2$. The eigenvalue
$\mu$ has the order of ${\rm O}(\epsilon)$. By continuity of
isolated eigenvalues ${\cal H}$ with respect to perturbation terms
of the Lyapunov--Schmidt reductions, we obtain the result
(\ref{eigenvalues-H-asymptotic}).

(iii) In the limit of $\epsilon = 0$, where $\epsilon = {\rm
O}(e^{-\kappa_0 L})$, the linearized operator $\partial_z {\cal
H}$ for the $n$-th two-pulse solution $\phi_n(z)$ has a
four-dimensional algebraic kernel according to the two-dimensional
kernel of the one-pulse solution (\ref{algebraic-kernel}). By the
translation invariance, the two-dimensional algebraic kernel
survives for small $\epsilon$ with the eigenfunctions
$\{\phi_n'(z), \partial_c \phi_n(z)\}$. Two eigenvalues $\lambda$
of the operator $\partial_z {\cal H}$ may bifurcate from the zero
eigenvalue at the order of ${\rm O}(\sqrt{\epsilon})$. The
splitting of the zero eigenvalue in the problem $\partial_z {\cal
H} v = \lambda v$ is considered by the perturbation theory,
\begin{equation}
\label{eigenvector-0-L} v(z) = -\alpha_1 \Phi'(z-s) - \alpha_2
\Phi'(z+s) + \beta_1
\partial_c \Phi(z-s) + \beta_2 \partial_c \Phi(z+s) + V(z),
\end{equation}
where $(\alpha_1,\alpha_2,\beta_1,\beta_2)$ are coordinates of the
projections to the algebraic kernel of $\partial_z {\cal H}$ at
$\epsilon = 0$ and $V(z)$ is the remainder term of the order of
${\rm O}(\epsilon)$. By projecting the eigenvalue problem
$\partial_z {\cal H} v = \lambda v$ to the algebraic kernel of the
adjoint operator $-{\cal H} \partial_z$ and neglecting the terms of
the order of ${\rm O}(\lambda \epsilon, \epsilon^2)$, we immediately
find at the order of ${\rm O}(\lambda)$ that $\beta_j = \lambda
\alpha_j$, $j = 1,2$, while $(\alpha_1,\alpha_2)$ satisfy a reduced
eigenvalue problem at the order of ${\rm O}(\lambda^2,\epsilon)$:
\begin{eqnarray*}
\frac{1}{2} \lambda^2 P'(c) \alpha_1 = -\tilde{W} \alpha_1 +
W''(L_n) \alpha_2, \qquad \frac{1}{2} \lambda^2 P'(c) \alpha_2 =
W''(L_n) \alpha_1 -\tilde{W} \alpha_2,
\end{eqnarray*}
where $P(c) = \| \Phi \|^2_{L^2}$ and $\tilde{W} = W''(L_n)$. The
non-zero eigenvalue $\lambda^2$ at the leading order is
$$
\lambda^2 = \frac{2 Q(c) \mu}{P'(c)} = - \frac{4W''(L_n)}{P'(c)}.
$$
Isolated eigenvalues $\partial_z {\cal H}$ are continuous with
respect to perturbation terms of the Lyapunov--Schmidt reductions,
so that we immediately obtain the result
(\ref{eigenvalues-asymptotic}) for $\lambda \in \mathbb{R}$ when
$W''(L_n) < 0$. In order to prove (\ref{eigenvalues-asymptotic})
for $\lambda \in i \mathbb{R}$ when $W''(L_n) > 0$, we compute
asymptotically the energy quadratic form
$$
({\cal H} v, v) = -4 W''(L_n) - P'(c) |\lambda|^2 + {\rm
O}(\lambda \epsilon,\epsilon^2),
$$
where $v(z)$ is given by the eigenfunction (\ref{eigenvector-0-L})
with $\alpha_1 = -\alpha_2 = 1$ and $\beta_j = \lambda \alpha_j$,
$j = 1,2$. When $\lambda \in i \mathbb{R}$ and $W''(L_n) > 0$, we
have $({\cal H} v, v) < 0$ up to the leading order, such that
$\lambda$ is an eigenvalue of negative Krein signature. Continuity
of the eigenvalues of negative Krein signature (even although the
eigenvalues $\lambda \in i \mathbb{R}$ are embedded into the
continuous spectrum of $\partial_z {\cal H}$) follows from
Pontryagin's Invariant Subspace Theorem (Theorem 1 in
\cite{ChPel}). By using the exponentially weighted spaces
\cite{PW94}, continuity of eigenvalues $\lambda \in i \mathbb{R}$
holds and the bound (\ref{eigenvalues-asymptotic}) is obtained for
an eigenvalue $\lambda \in \mathbb{C}$ in a neighborhood of the
point $\lambda \in i \mathbb{R}$. Note that the continuity of one
pair of eigenvalues $\lambda \in \mathbb{C}$ satisfying
(\ref{eigenvalues-asymptotic}) does not exclude the possibility of
existence of another pair of eigenvalues $\lambda \in \mathbb{C}$
(in the exponentially weighted space with a negative weight) which
also satisfies (\ref{eigenvalues-asymptotic}).
\end{proof}

\begin{remark}
{\rm Theorem \ref{theorem-slow-dynamics} is a modification of
Theorems 1 and 2 in \cite{S98} (see also \cite{Lin90}) in
applications to the ODE (\ref{ODE}) and the spectral problem
(\ref{stability-problem}). The proof is lighten compared to the
original proofs in \cite{Lin90,S98}.  We note that the persistence
of purely imaginary eigenvalues (\ref{eigenvalues-asymptotic}) in
the full problem (\ref{stability-problem}) cannot be proved with
the Lyapunov--Schmidt reduction method since the essential
spectrum of $\partial_z {\cal H}$ occurs on the imaginary axis
(contrary to the standard assumption of Theorem 2 in \cite{S98}
that the essential spectrum occurs in the left half-plane.)}
\end{remark}

\begin{remark}
{\rm The following claim from the Gorshkov--Ostrovsky perturbative
procedure \cite{GO81,GOP84} illustrates the role of $W(L)$ as the
effective interaction potential for the slow dynamics of a
two-pulse solution:

\noindent {\em {\bf Claim:} Let $C_1, C_2$ be some positive
constants. For the initial time interval $0 \leq t \leq C_1
e^{\kappa_0 L/2}$ and up to the leading order ${\rm
O}(e^{-\kappa_0 L})$, the two-pulse solutions of the fifth-order
KdV equation (\ref{kdv}) can be written as the decomposition
$$
u(x,t) = \Phi(x - ct - s(t)) + \Phi(x - ct + s(t)) + U(x,t),
$$
where $\| U \|_{L^{\infty}} \leq C_2 e^{-\kappa_0 L}$ and the slow
dynamics of $L(t) = 2 s(t)$ is represented by the Newton's particle
law:
\begin{equation}
\label{Newton-law} P'(c) \ddot{L} = - 4 W'(L).
\end{equation}
}

\noindent Although rigorous bounds on the time interval and the
truncation error of the Newton's particle law were recently found
in the context of NLS solitons in external potentials (see
\cite{Sigal} and references therein), the above claim was not
proved yet in the context of two-pulse solutions of the
fifth-order KdV equation (\ref{kdv}). We note that perturbation
analysis that leads to the Newton's particle law
(\ref{Newton-law}) cannot be used to claim persistence and
topological equivalence of dynamics of the Newton's particle to
the full dynamics of two-pulse solutions.}
\end{remark}

According to Theorem \ref{theorem-slow-dynamics}, an infinite set
of extrema of $W(L)$  generates a sequence of equilibrium
configurations for the two-pulse solutions in Theorem
\ref{theorem-two-pulses}. Since $P'(c) > 0$ by Theorem
\ref{theorem-one-pulse}(iii), the maxima points of $W(L)$
correspond to a pair of real eigenvalues $\lambda$ of the spectral
problem (\ref{stability-problem}), while the minima points of
$W(L)$ correspond to a pair of purely imaginary eigenvalues
$\lambda$. The two-pulse solutions at the maxima points are thus
expected to be linearly and nonlinearly unstable. The two-pulse
solutions at the minima points are stable within the leading-order
approximation (\ref{eigenvalues-asymptotic}) and within the
Newton's particle law (\ref{Newton-law}) (a particle with the
coordinate $L(t)$ performs a periodic oscillation in the potential
well). Correspondence of these predictions to the original PDE
(\ref{kdv}) is a subject of the present article. We will compute
the interaction potential $W(L)$ and the sequence of its extrema
points $\{L_n\}_{n \in \mathbb{N}}$, as well as the numerical
approximations of the two-pulse solutions of the ODE (\ref{ODE})
and of the eigenvalues of the operator $\partial_z {\cal H}$ in
(\ref{stability-problem}).

\section{Iterations of the Petviashvili's method for two-pulse
solutions}

We address the Petviashvili's iteration method for numerical
approximations of solutions of the fourth-order ODE (\ref{ODE})
with $c > 0$. See review of literature on the Petviashvili's
method in \cite{PelStep}. By using the standard Fourier transform
in $L^1(\mathbb{R}) \cap L^2(\mathbb{R})$
$$
\hat{\phi}(k) = \int_{\mathbb{R}} \phi(z) e^{-ikz} dz, \qquad k \in
\mathbb{R},
$$
we reformulate the ODE (\ref{ODE}) as a fixed-point problem in the
Sobolev space $H^2(\mathbb{R})$:
\begin{equation}
\label{fixed-point} \hat{\phi}(k) = \frac{\widehat{\phi^2}(k)}{(c +
k^2 + k^4)}, \qquad k \in \mathbb{R},
\end{equation}
where $\widehat{\phi^2}(k)$ can be represented by the convolution
integral of $\hat{\phi}(k)$ to itself. An even real-valued
solution $\phi(-z) = \phi(z)$ of the ODE (\ref{ODE}) in
$H^2(\mathbb{R})$ is equivalent to the even real-valued solution
$\hat{\phi}(-k) = \hat{\phi}(k)$ of the fixed-point problem
(\ref{fixed-point}). Let us denote the space of all even functions
in $H^2(\mathbb{R})$ by $H^2_{\rm ev}(\mathbb{R})$ and consider
solutions of the fixed-point problem (\ref{fixed-point}) in
$H^2_{\rm ev}(\mathbb{R})$.

Let $\{ \hat{u}_n(k) \}_{n = 0}^{\infty}$ be a sequence of Fourier
transforms in $H_{\rm ev}^2(\mathbb{R})$ defined recursively by
\begin{equation}
\label{Petviashvili}
 \hat{u}_{n+1}(k) = M^2_{n} \frac{\widehat{u_{n}^{2}}(k)}{(c + k^2 + k^4)},
\end{equation}
where $\hat{u}_0(k) \in H_{\rm ev}^2(\mathbb{R})$ is a starting
approximation and $M_n \equiv M[\hat{u}_n]$ is the Petviashvili's
stabilizing factor defined by
\begin{equation} \label{factor}
M[\hat{u}] = \frac{\int_{\mathbb{R}} (c + k^2 + k^4) \left[
\hat{u}(k)\right]^2 dk}{ \int_{\mathbb{R}} \hat{u}(k)
\widehat{u^{2}}(k)dk}.
\end{equation}
It follows from the fixed-point problem (\ref{fixed-point}) that
$M[\hat{\phi}] = 1$ for any solution $\hat{\phi} \in H^2_{\rm
ev}(\mathbb{R})$. The following theorem was proved in \cite{PelStep}
and reviewed in \cite{DemSchlag}.

\begin{theorem}
Let $\hat{\phi}(k)$ be a solution of the fixed-point problem
(\ref{fixed-point}) in $H_{\rm ev}^2(\mathbb{R})$. Let ${\cal H}$ be
the Jacobian operator (\ref{Jacobian}) evaluated at the
corresponding solution $\phi(z)$ of the ODE (\ref{ODE}). If ${\cal
H}$ has exactly one negative eigenvalue and a simple zero eigenvalue
and if
\begin{equation}
\label{condition-Petviashvili} {\rm either} \quad \phi(z) \geq 0
\qquad {\rm or} \quad \left| \inf_{z \in \mathbb{R}} \phi(z) \right|
< \frac{c}{2},
\end{equation}
then there exists an open neighborhood of $\hat{\phi}$ in
$H^2_{\rm ev}(\mathbb{R})$, in which $\hat{\phi}$ is the unique
fixed point and the sequence of iterations $\{ \hat{u}_n(k)
\}_{n=0}^{\infty}$ in (\ref{Petviashvili})--(\ref{factor})
converges to $\hat{\phi}$. \label{theorem-Petviashvili}
\end{theorem}

\begin{proof}
We review the basic steps of the proof, which is based on the
contraction mapping principle in a local neighborhood of
$\hat{\phi}$ in $H^2_{\rm ev}(\mathbb{R})$. The linearization of
the iteration map (\ref{Petviashvili}) at the solution $\phi$ is
rewritten in the physical space $z \in \mathbb{R}$ as follows:
\begin{equation}
\label{linearized-iteration} v_{n+1}(z) = - 2 \alpha_n \phi(z) +
v_n(z) - (c - \partial_z^2 + \partial_z^4)^{-1} {\cal H} v_n(z),
\end{equation}
where $\alpha_n$ is a projection of $v_n$ onto $\phi^2$ in
$L^2(\mathbb{R})$:
$$
\alpha_n = \frac{(\phi^2,v_n)}{(\phi^2,\phi)},
$$
such that $u_n = \phi + v_n$ and $M_n = 1 - \alpha_n$ to the linear
order. The operator ${\cal T} = (c - \partial_z^2 +
\partial_z^4)^{-1} {\cal H}$ is a self-adjoint operator in
Pontryagin space $\Pi_0$ defined by the inner product
$$
\forall f,g \in \Pi_0 : \quad [f,g] = ((c - \partial_z^2 +
\partial_z^4) f, g).
$$
See \cite{ChPel} for review of Pontryagin spaces and the
Pontryagin Invariant Subspace Theorem. Since $c > 0$, the
Pontryagin space $\Pi_0$ has zero index and, by the Pontryagin
Theorem, the operator ${\cal T}$ in $\Pi_0$ has exactly one
negative eigenvalue, a simple kernel and infinitely many positive
eigenvalues. The eigenfunctions for the negative and zero
eigenvalues are known exactly as
$$
{\cal T} \phi = - \phi, \qquad {\cal T} \phi'(z) = 0.
$$
Due to orthogonality of the eigenfunctions in the Pontryagin space
$\Pi_0$ and the relation
$$
\phi^2 = (c - \partial_z^2 + \partial_z^4) \phi,
$$
we observe that $\alpha_n$ is a projection of $v_n$ to $\phi$ in
$\Pi_0$, which satisfies the trivial iteration map:
$$
\alpha_{n+1} = 0, \qquad n \geq 1.
$$
Projection of $v_n$ to $\phi'$ in $\Pi_0$ is zero since $v_n \in
H^2_{\rm ev}(\mathbb{R})$. As a result, the linearized iteration
map (\ref{linearized-iteration}) defines a contraction map if the
maximal positive eigenvalue of ${\cal T}$ in $L^2(\mathbb{R})$ is
smaller than $2$. However,
\begin{equation}
\sigma\left({\cal T} \biggr|_{L^2} \right) - 1 \leq -2
\inf_{\|u\|_{L^2} = 1} \left( u,(c -
\partial_z^2 + \partial_z^4)^{-1} \phi(z) u \right).
\label{max-inequality}
\end{equation}
If $\phi(z) \geq 0$ on $z \in \mathbb{R}$, the right-hand-side of
(\ref{max-inequality}) is zero. Otherwise, the right-hand-side of
(\ref{max-inequality}) is bounded from above by $\frac{2}{c} \left|
\inf_{z \in \mathbb{R}} \phi(z) \right|$, which leads to the
condition (\ref{condition-Petviashvili}).
\end{proof}

\begin{corollary}
Let $\phi(z)$ be a one-pulse solution of the ODE (\ref{ODE}) with $c
> 0$ defined by Theorem \ref{theorem-one-pulse}. Then, the iteration method
(\ref{Petviashvili})--(\ref{factor}) converges to $\phi(z)$ in a
local neighborhood of $\phi$ in $H^2_{\rm ev}(\mathbb{R})$ provided
that the condition (\ref{condition-Petviashvili}) is met.
\label{corollary-Petviashvili}
\end{corollary}

The condition (\ref{condition-Petviashvili}) is satisfied for the
positive exact solution (\ref{nexact}) for $c = \frac{36}{169}$.
Since the one-pulse solution is positive definite for $0 < c <
\frac{1}{4}$ \cite{AT91}, it is also satisfied for all values of
$c \in (0,\frac{1}{4})$. However, the solution is sign-indefinite
for $c \geq \frac{1}{4}$, such that the condition
(\ref{condition-Petviashvili}) must be checked {\em a posteriori},
after a numerical approximation of the solution is obtained.

Besides the convergence criterion described in Theorem
\ref{theorem-Petviashvili}, there are additional factors in the
numerical approximation of the one-pulse solution of the ODE
(\ref{ODE}) which comes from the discretization of the Fourier
transform, truncation of the resulting Fourier series, and
termination of iterations within the given tolerance bound. These
three numerical factors are accounted by three numerical
parameters:

\begin{itemize}
\item[(i)] $d$ - the half-period of the computational interval $z
\in [-d,d]$ where the solution $\phi(z)$ is represented by the
Fourier series for periodic functions;

\item[(ii)] $N$ - the number of terms in the partial sum for the
truncated Fourier series such that the grid size $h$ of the
discretization is $h = 2d/N$;

\item[(iii)] $\varepsilon$ - the small tolerance distance that measures deviation of $M_n$ from $1$ and
the distance between two successive approximations, such that the
method can be terminated at the iteration $n$ if
$$
E_M \equiv | M_n - 1 | < \varepsilon \qquad \mbox{and} \qquad
E_{\infty} \equiv \| u_{n+1} - u_n \|_{L^{\infty}} < \varepsilon.
$$
and $\tilde{\phi} = u_n(z)$ can be taken as the numerical
approximation of the solution $\phi(z)$.
\end{itemize}

The numerical approximation depends weakly of the three numerical
parameters, provided (i) $d$ is much larger than the half-width of
the one-pulse solution, (ii) $N$ is sufficiently large for
convergence of the Fourier series, and (iii) $\varepsilon$ is
sufficiently small above the level of the round-off error. Indeed,
the constraint (i) ensures that the truncation error is
exponentially small when the one-pulse solution is replaced by the
periodic sequence of one-pulse solutions in the trigonometric
approximation \cite{Scheel}. The constraint (ii) ensures that the
remainder of the Fourier partial sum is smaller than any inverse
power of $N$ (by Theorem \ref{theorem-one-pulse}(i), all
derivatives of the function $\phi(z)$ are continuous)
\cite{Trefethen}. The constraint (iii) specifies the level of
accuracy achieved when the iterations of the method
(\ref{Petviashvili})--(\ref{factor}) are terminated. While we do
not proceed with formal analysis of the three numerical factors
(see \cite{DemSchlag} for an example of this analysis), we
illustrate the weak dependence of three numerical factors on the
example of the numerical approximation $\tilde{\phi}(z)$ of the
exact one-pulse solution (\ref{nexact}), which exists for $c =
\frac{36}{169}$. Numerical implementation of the iteration method
(\ref{Petviashvili})--(\ref{factor}) was performed in MATLAB
according to a standard toolbox of the spectral methods
\cite{Trefethen}.

Figure 1 displays the distance $E = \| \tilde{\phi} - \phi
\|_{L^{\infty}}$ versus the three numerical factors $d$, $h$, and
$\varepsilon$ described above. The left panel shows that the error
$E$ converges to  the numerical zero, which is ${\rm O}(10^{-15})$
in MATLAB under the Windows platform, when the step size $h$ is
reduced, while $d = 50$ and $\varepsilon = 10^{-15}$ are fixed.
The middle panel computed for $h = 1$ and $\varepsilon = 10^{-15}$
shows that the error $E$ converges to the level ${\rm
O}(10^{-13})$ when the half-width $d$ is enlarged. The numerical
zero is not reached in this case because the step size $h$ is not
sufficiently small. The right panel computed for $h = 1$ and $d =
50$ shows that the error $E$ converges to the same level ${\rm
O}(10^{-13})$ as the tolerance bound $\varepsilon$ is reduced. In
all approximations that follow, we will specify $h = 0.01$, $d =
50$ and $\varepsilon = 10^{-15}$ to ensure that the error of the
iteration method (\ref{Petviashvili})--(\ref{factor}) for
one-pulse solutions is on the level of the numerical zero ${\rm
O}(10^{-15})$.

Figure 2 (left) shows the numerical approximation of the one-pulse
solutions for $c = 4$, where the small-amplitude oscillations of
the exponentially decaying tail are visible. We check a posteriori
the condition (\ref{condition-Petviashvili}) for non-positive
one-pulse solutions $\left| \inf_{z \in \mathbb{R}} \phi(z)
\right| < 2$ for $c = 4$. Figure 2 (right) displays convergence of
the errors $E_M = |M_n - 1|$ and $E_{\infty} = \| u_{n+1} - u_n
\|_{L^{\infty}}$ computed dynamically at each $n$ as $n$
increases. We can see that the error $E_M$ converges to zero much
faster than the error $E_{\infty}$, in agreement with the
decomposition of the linearized iterative map
(\ref{linearized-iteration}) into the one-dimensional projection
$\alpha_n$ and the infinite-dimensional orthogonal compliment (see
the proof of Theorem \ref{theorem-Petviashvili}). In all further
approximations, we will use the error $E_{\infty}$ for termination
of iterations and detecting its minimal values since $E_{\infty}$
is more sensitive compared to $E_M$.

\begin{figure}
\begin{center}
\includegraphics [width=14cm] {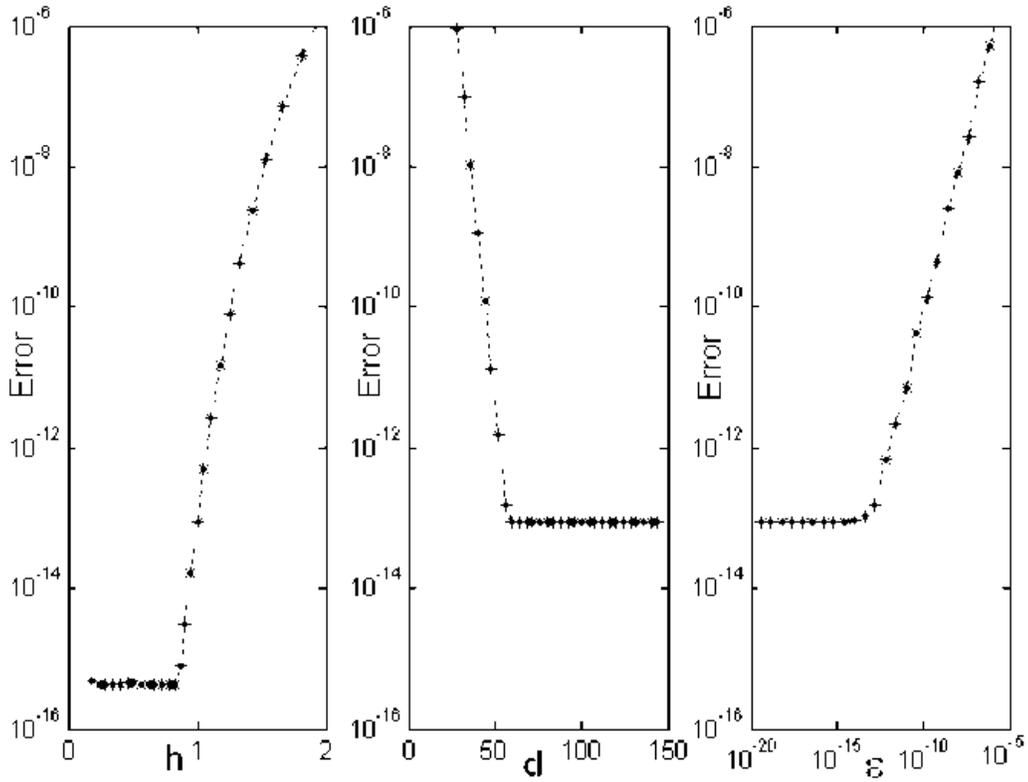}
\end{center}
\caption{The distance $E = \| \tilde{\phi} - \phi
\|_{L^{\infty}}$ for the ODE (\ref{ODE}) with $c = \frac{36}{169}$
versus the half-period $d$ of the computational interval, the step
size $h$ of the discretization, and the tolerance bound
$\varepsilon$.}
\end{figure}

\begin{figure}
\begin{center}
\includegraphics [width=4.8cm] {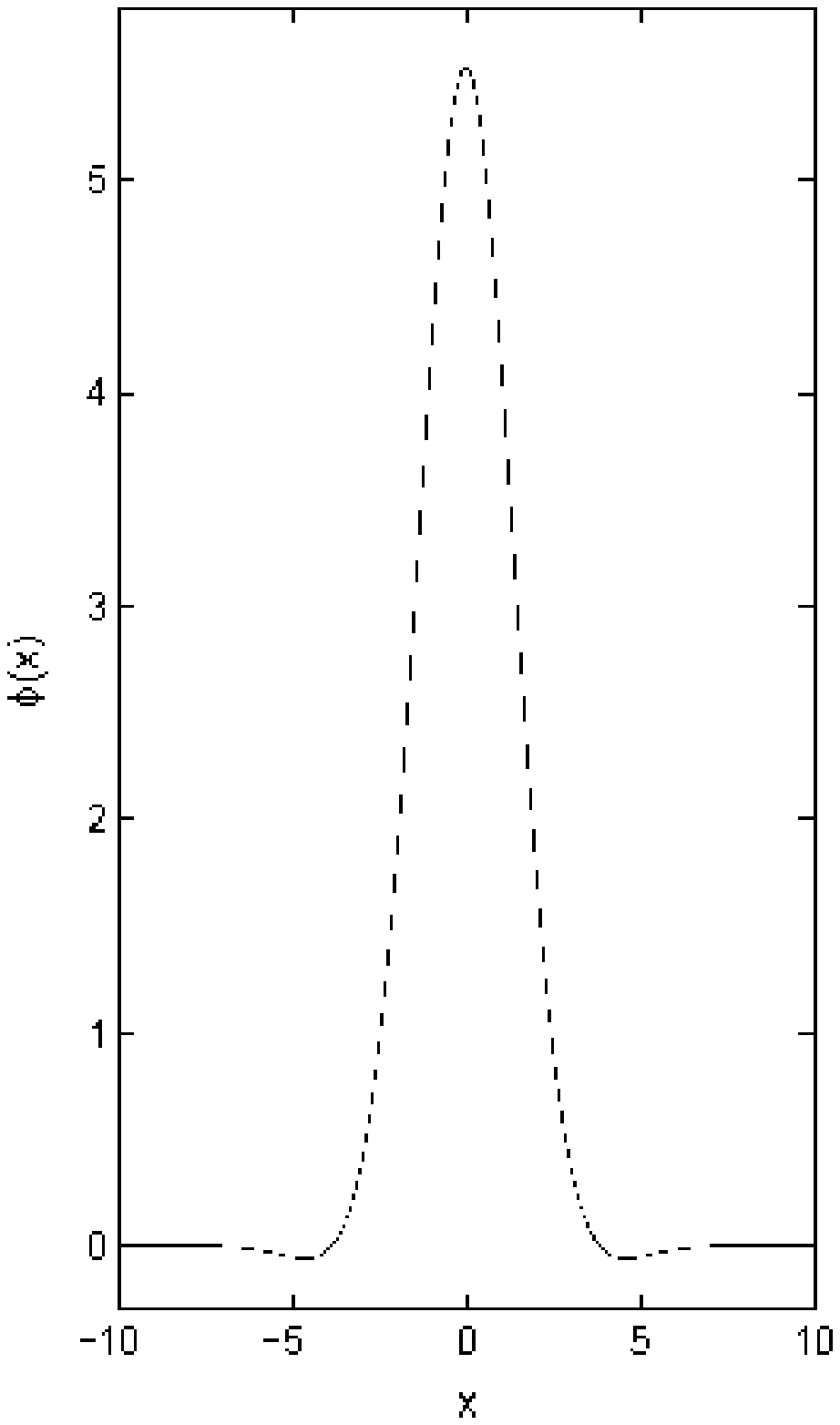}
\includegraphics [width=4.5cm] {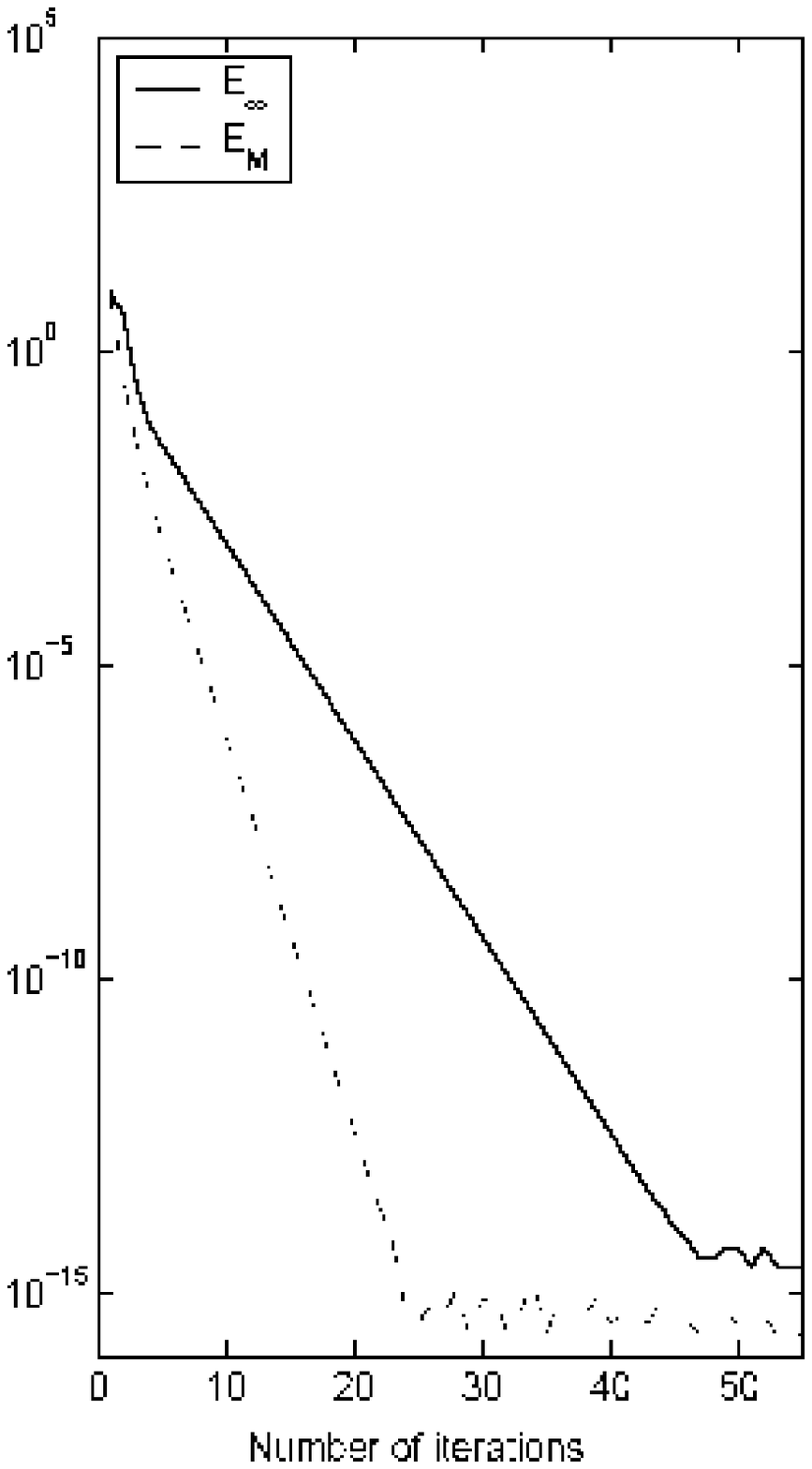}
\end{center}
\caption{One-pulse solutions of the ODE (\ref{ODE}) with $c = 4$
(left) and convergence of the errors $E_M$ and $E_{\infty}$ to zero
versus the number of iterations $n$.}
\end{figure}

Since the numerical approximations $\tilde{\phi}(z)$ of one-pulse
solutions can be computed for any value of $c > 0$, one can use
$\tilde{\phi}(z)$ for a given $c$ and compute the effective
interaction potential (\ref{effective-potential}), which defines
the extremal values $\{ L_n \}_{n \in \mathbb{N}}$. Theorem
\ref{theorem-slow-dynamics} guarantees that the two-pulse solution
$\phi_n(z)$ consists of two copies of the one-pulse solutions
separated by the distance $L$ near the point $L_n$ where $W'(L_n)
= 0$ and $W''(L_n) \neq 0$. Table 1 shows the first four values of
the sequence $\{ L_n \}_{n=1}^{\infty}$ for $c = 1$ (where $s_n =
L_n/2$ is the half-distance between the pulses). It also shows the
corresponding values from the first four numerical approximations
of two-pulse solutions $\phi_n(z)$ (obtained below) and the
computational error computed from the difference of the two
numerical approximations. We can see that the error decreases for
larger indices $n$ in the sequence $\{ L_n \}_{n \in \mathbb{N}}$
since the Lyapunov--Schmidt reductions of Theorem
\ref{theorem-slow-dynamics} become more and more accurate in this
limit.

\begin{center}
\begin{tabular}{|l|l|l|l|}
    \hline two-pulse solution & effective potential & root finding & error  \\
    \hline
$s = s_1$   & $5.058733328146916$   & $5.079717398028492$  & $0.02098406988158$   \\
$s = s_2$   & $8.196800619090793$   & $8.196620796452045$  & $1.798226387474955 \cdot 10^{-4}$   \\
$s = s_3$   & $11.338414567609066$ & $11.338406246900558$  & $8.320708507980612 \cdot 10^{-6}$   \\
$s = s_4$   & $14.479997655627219$ & $14.479996635578457$  & $1.020048761901649 \cdot 10^{-6}$  \\
    \hline
\end{tabular}
\end{center}

{\bf Table 1:} The first four members of the sequence of two-pulse
solutions of the ODE (\ref{ODE}) with $c = 1$.

\vspace{1cm}

By Theorem \ref{theorem-slow-dynamics}(ii), the Jacobian operator
${\cal H}$ associated with a two-pulse solution $\phi(z)$ has one
finite negative eigenvalue in the space of even functions
$H^2_{\rm ev}(\mathbb{R})$ and one small eigenvalue which is
either negative or positive depending on the sign of $W''(L_n)$.
This small eigenvalue leads to either weak divergence or weak
convergence of the Petviashvili method in a local neighborhood of
$\phi$ in $H^2_{\rm ev}(\mathbb{R})$. Even if the small eigenvalue
is positive and the algorithm is weakly convergent, the truncation
error from the numerical discretization may push the small
eigenvalue to a negative value and lead thus to weak divergence of
the iterations.

Figure 3 illustrates typical behaviors of the errors $E_M$ and
$E_{\infty}$ versus $n$ for the starting approximation
\begin{equation}
\label{starting_approximation} u_0(z) = U_0(z-s) + U_0(z+s),
\end{equation}
where $U_0(z)$ is a starting approximation of a sequence
$\{u_n(z)\}_{n \in \mathbb{N}}$ which converges to the one-pulse
solution $\Phi(z)$ and $s$ is a parameter defined near $L_n/2$ for
the $n$-th two-pulse solution $\phi_n(z)$. The left panel shows
iterations for $s$ near $s_1$ and the right panel shows iterations
for $s$ near $s_2$. Since $W''(L_1) > 0$ and $W''(L_2) < 0$, the
two-pulse solution $\phi_1(z)$ leads to the weak divergence of the
iteration method (\ref{Petviashvili})--(\ref{factor}), while the
two-pulse solution $\phi_2(z)$ leads to the weak convergence of
the method.

At the initial stage of iterations, both errors $E_M$ and
$E_{\infty}$ quickly drops to small values, since the starting
iterations $U_0(z\mp s)$ converge to the one-pulse solutions
$\Phi(z\mp s)$ while the contribution from the overlapping tails
of $\Phi(z \mp s)$ is negligible. However, at the later stage of
iterations, both errors either start to grow (the left panel of
Figure 3) or stop to decrease (the right panel). As it is
explained above, this phenomenon is related to the presence of
zero eigenvalue of ${\cal H}$ in $H^2_{\rm ev}(\mathbb{R})$ which
bifurcates to either positive or negative values due to
overlapping tails of $\Phi(z \mp s)$ and due to the truncation
error. At the final stage of iterations on the left panel of
Figure 3, the numerical approximation $u_n(z)$ converges to the
one-pulse solution $\Phi(z)$ centered at $z = 0$ and both errors
quickly drop to the numerical zero, which occurs similarly to the
right panel of Figure 2. No transformation of the solution shape
occurs for large $n$ on the right panel of Figure 3.

The following theorem defines an effective numerical algorithm,
which enables us to compute the two-pulse solutions from the weakly
divergent iterations of the Petviashvili's method
(\ref{Petviashvili})--(\ref{factor}).

\begin{theorem}
Let $\phi_n(z)$ be the $n$-th two-pulse solution of the ODE
(\ref{ODE}) defined by Theorems \ref{theorem-two-pulses} and
\ref{theorem-slow-dynamics}. There exists $s = s_*$ near $s =
L_n/2$ such that the iteration method
(\ref{Petviashvili})--(\ref{factor}) with the starting
approximation $u_0(z) = \Phi(z-s) + \Phi(z+s)$ converges to
$\phi_n(z)$ in a local neighborhood of $\phi_n$ in $H^2_{\rm
ev}(\mathbb{R})$. \label{theorem-convergence-two-pulse}
\end{theorem}

\begin{proof}
The iteration operator (\ref{Petviashvili})--(\ref{factor}) in the
neighborhood of the two-pulse solution $\phi_n$ in $H^2_{\rm
ev}(\mathbb{R})$ can be represented into an abstract form
$$
v_{n+1} = L(\epsilon) v_n + N(v_n,\epsilon), \qquad n \in
\mathbb{N},
$$
where $L(\epsilon)$ has a unit eigenvalue at $\epsilon = 0$,
$N(v_n,\epsilon)$ is $C^k$ in $v_n \in H^2_{\rm ev}$ with $k \geq
2$ such that $N(0,0) = D_v N(0,0) = 0$, $v_n$ is a perturbation of
$u_n$ to the fixed point $\phi_n$, and $\epsilon$ is a small
parameter for two-pulse solutions defined in Theorem
\ref{theorem-slow-dynamics}. By the Center Manifold Reduction for
quasi-linear discrete systems (Theorem 1 in \cite{James03}), there
exists a one-dimensional smooth center manifold in a local
neighborhood of $\phi_n$ in $H^2_{\rm ev}(\mathbb{R})$. Let $\xi$
be a coordinate of the center manifold such that $\xi \in
\mathbb{R}$, $\xi = 0$ corresponds to $v = 0$, and the dynamics on
the center manifold is
$$
\xi_{n+1} = \mu(\epsilon) \xi_n + f(\xi_n,\epsilon), \qquad n \in
\mathbb{N},
$$
where $\mu(\epsilon)$ with $\mu(0) = 1$ is an eigenvalue of the
linearized operator ${\cal T}(\epsilon)$ at $\phi_n$ in $H^2_{\rm
ev}(\mathbb{R})$ defined in Theorem \ref{theorem-Petviashvili} and
$f(\xi_n,\epsilon)$ is $C^k$ in $\xi \in \mathbb{R}$ with $k \geq
2$ and $f(0,0) = \partial_{\xi} f(0,0) = 0$. Consider the
one-parameter starting approximation $u_0(z) = \Phi(z-s) +
\Phi(z+s)$ in a neighborhood of $\phi_n$ in $H^2_{\rm
ev}(\mathbb{R})$, where $s$ is close to the value $s = s_n$
defined in Theorem \ref{theorem-slow-dynamics}. By the time
evolution of the hyperbolic component of $v_n$ (see Lemma 2 in
\cite{James03}), the sequence $v_n$ approaches to the center
manifold with the coordinate $\xi_n$. Iterations of $\xi_n$ are
sign-definite in a neighborhood of $\xi = 0$. Moreover, there
exists $s_1 < s_n$ and $s_2 > s_n$, such that the sequences
$\{\xi_n(s_1)\}_{n \in \mathbb{N}}$ and $\{\xi_n(s_2)\}_{n \in
\mathbb{N}}$ are of opposite signs. By smoothness of $v_n$ and
$\xi_n$ from parameter $s$, there exists a root $s_*$ in between
$s_1 < s_* < s_2$ such that $\xi_n(s_*) = 0$ for all $n \in
\mathbb{N}$.
\end{proof}

\begin{remark}
{\rm The proof of Theorem \ref{theorem-convergence-two-pulse} does
not require that the root $s_*$ be unique for the one-parameter
starting approximation $u_0(z) = \Phi(z-s) + \Phi(z+s)$. Our
numerical computations starting with a more general approximation
(\ref{starting_approximation}) show that the root $s_*$ is unique
in a neighborhood of $s_n$. }
\end{remark}

In order to capture the two-pulse solutions according to Theorem
\ref{theorem-convergence-two-pulse}, we compute the minimum of the
error $E_{\infty}$ for different values of $s$ and find numerically
a root $s = s_*$ of the function
$$
f(s) = \min_{0 \leq n \leq n_0}(E_{\infty}),
$$
where $n_0$ is the first iterations after which the value of
$E_{\rm infty}$ increases (in case of the left panel of Figure 3)
or remains unchanged (in case of the right panel of Figure 3). The
numerical root $s = s_*$ is found by using the secant method:
\begin{equation}
\label{secant-method} s_k = \frac{s_{k-2}f(s_{k-1}) -
s_{k-1}f(s_{k-2})}{f(s_{k-1}) - f(s_{k-2})}.
\end{equation}
The Petviashvili's method (\ref{Petviashvili})--(\ref{factor})
with the starting approximation (\ref{starting_approximation})
where $s$ is close to the root $s = s_*$ near the point $s = s_n$
converges to the two-pulse solution $\phi_n(z)$ within the
accuracy of the round-off error.

Figure 4 shows the graph of $f(s)$ near the value $s = s_1$ for $c
= 1$. (The graph of $f(s)$ near $s = s_2$ as well as other values
of $s_n$ look similar to Figure 4.) The left panel shows
uniqueness of the root, while the right panel shows the linear
behavior of $f(s)$ near $s = s_*$ which indicates that the root is
simple. Numerical approximations for the first four values of the
sequence $\{s_n\}_{n \in \mathbb{N}}$ obtained in this root
finding algorithm are shown in Table 1. We note that the number of
iterations $N_{\rm h}$ of the secant method (\ref{secant-method})
decreases with larger values of $n$, such that $N_{\rm h} = 14$
for $n = 1$, $N_{\rm h} = 12$ for $n = 2$, $N_{\rm h} = 10$ for $n
= 3$ and $N_{\rm h} = 9$ for $n = 4$, while the number of
iterations of the Petviashili's method for each computation does
not exceed $100$ iterations.

Figure 5 shows numerical approximations of the two-pulse solutions
for $c = 1$ and $c = 4$. We can see from the right panel that
two-pulse solutions with $c = 4$ resemble the two copies of the
one-pulse solutions from the left panel of Figure 2, separated by
the small-amplitude oscillatory tails.

\begin{figure}
\begin{center}
\includegraphics [width=5cm] {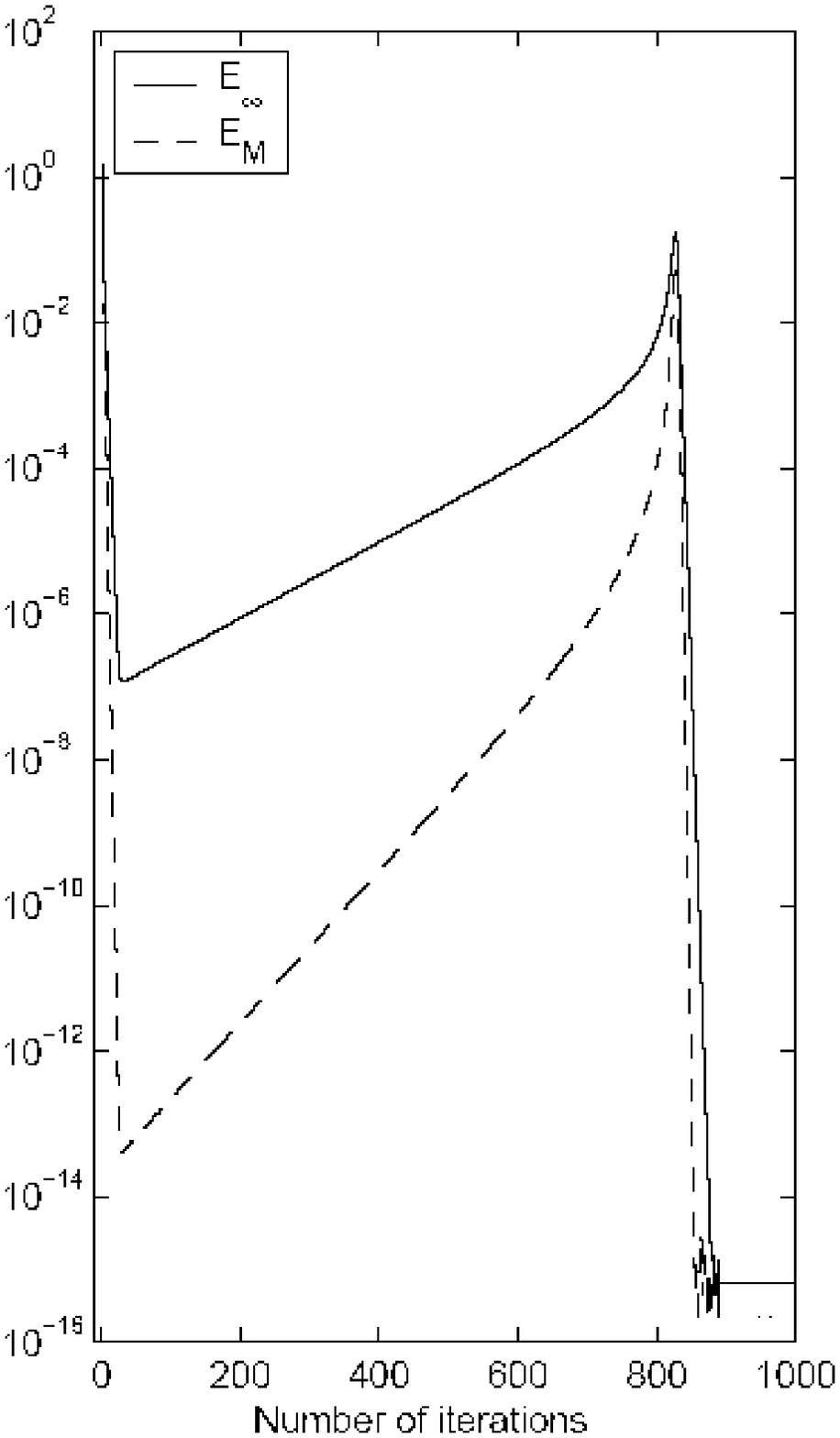}
\includegraphics [width=5cm] {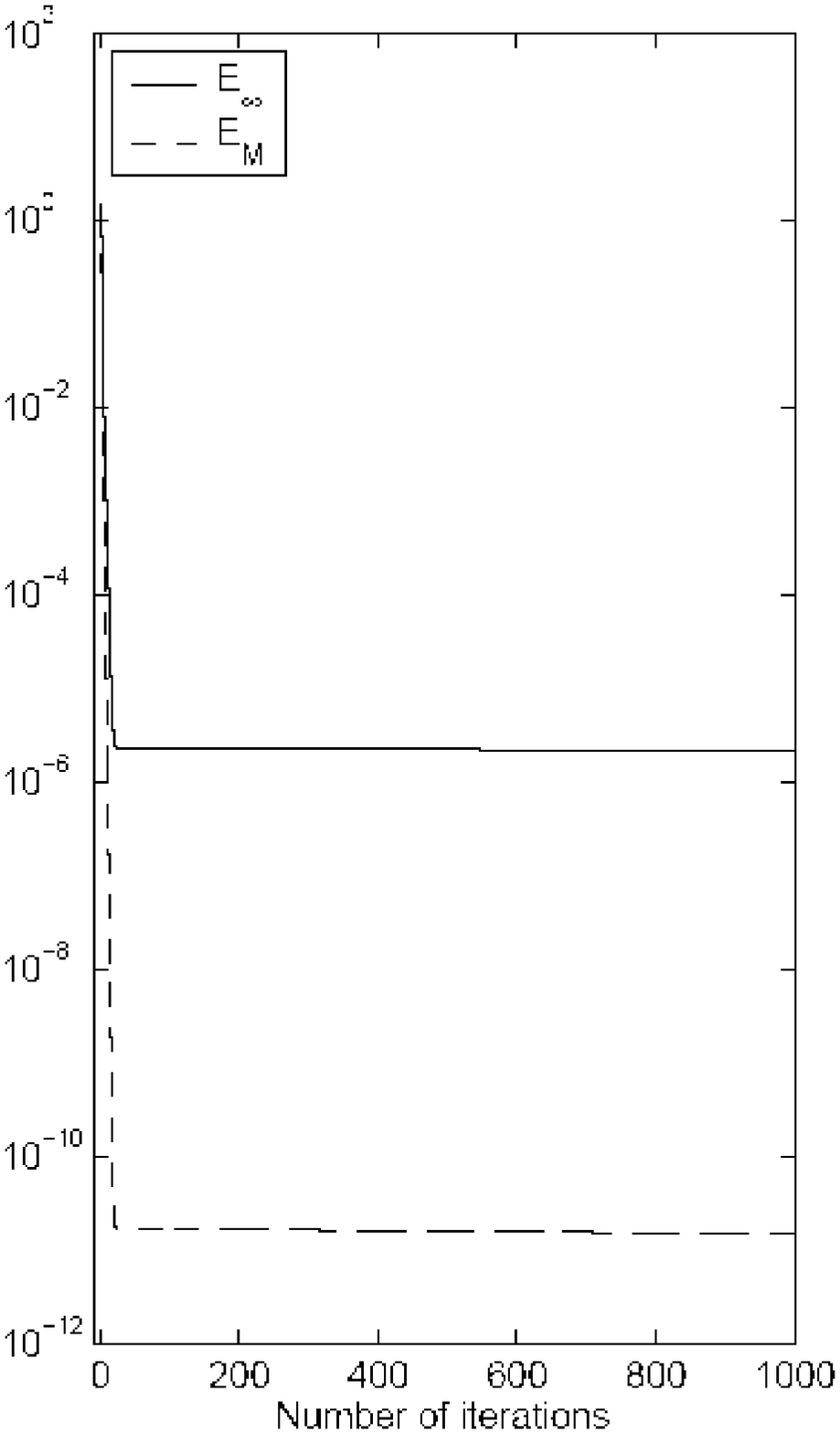}
\end{center}
\caption{Errors $E_M$ and $E_{\infty}$ versus the number of
iterations $n$ for the starting approximation
(\ref{starting_approximation}) with $s = 5.079$ (left panel) and
$s = 8.190$ (right panel). The other parameters are: $c = 1$, $d =
50$, $h = 0.01$ and $\varepsilon = 10^{-15}$.}
\end{figure}

\begin{figure}
\begin{center}
\includegraphics [width=5.6cm] {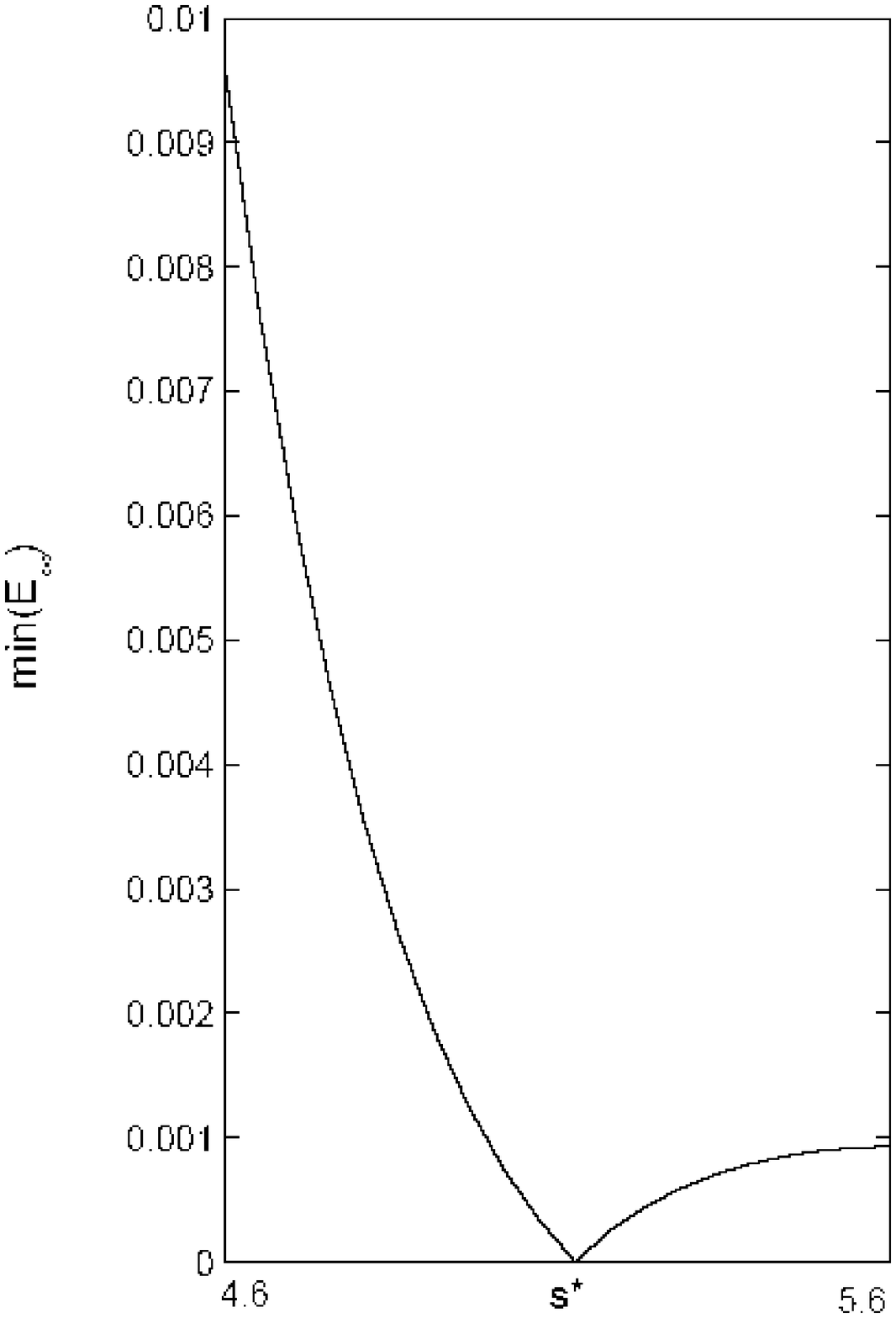}
\includegraphics [width=5cm] {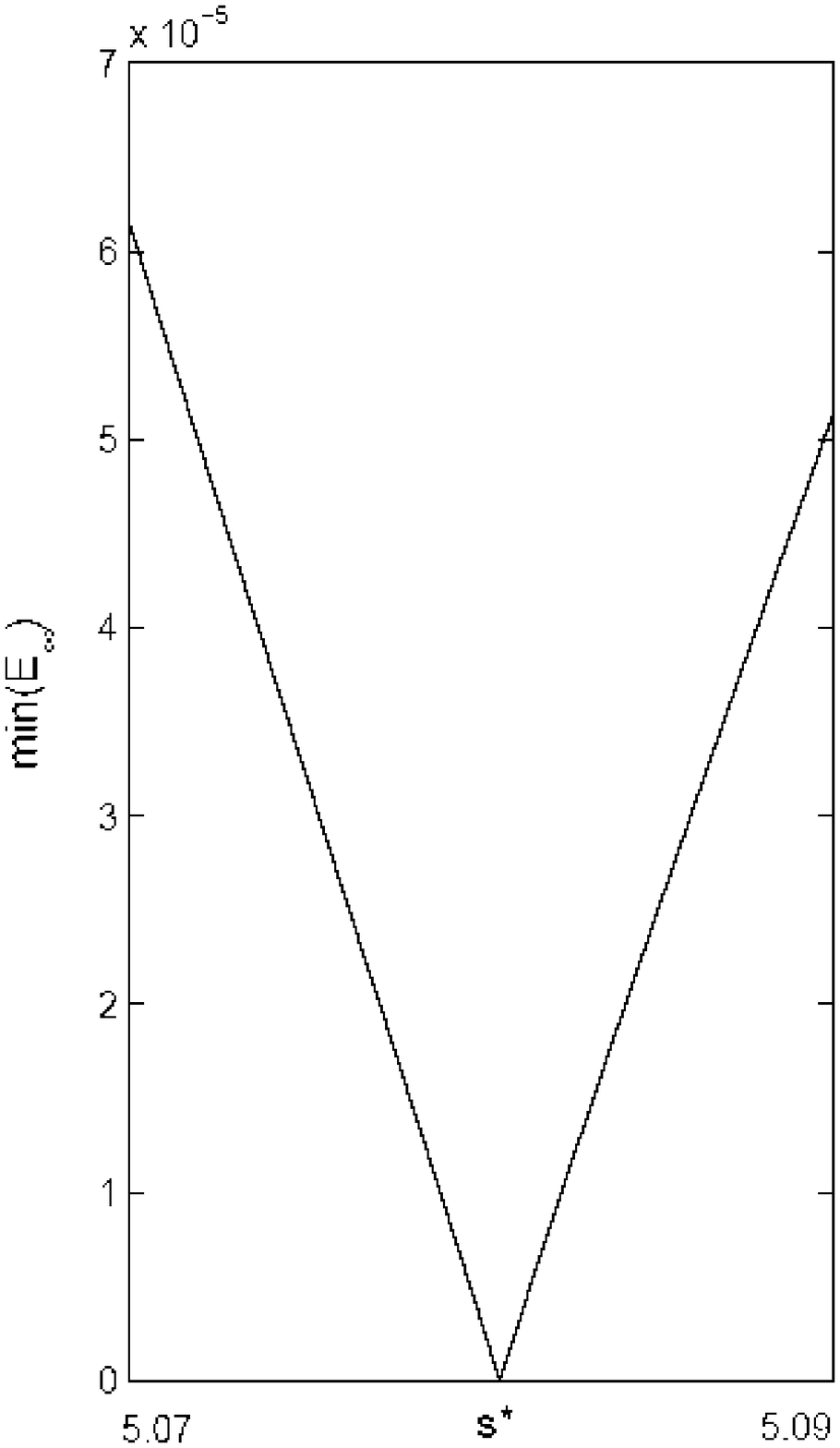}
\end{center}
\caption{Minimal value of $E_\infty$ versus $s$ near $s_1 = 5.080$
(left panel) and the zoom of the graph, which shows the linear
behavior of $f(s)$ near the root (right panel).}
\end{figure}

\begin{figure}
\begin{center}
\includegraphics [width=5.15cm] {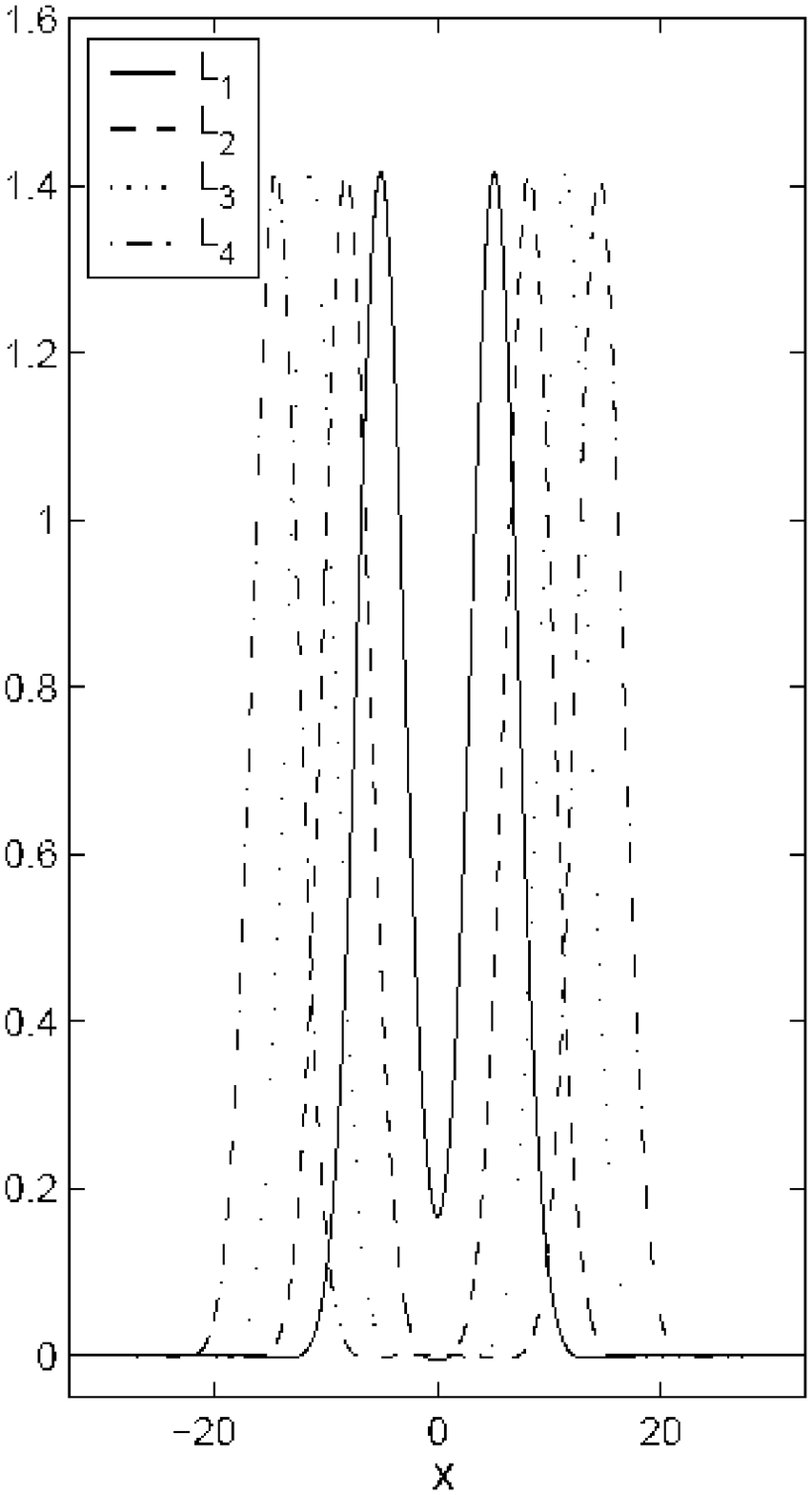}
\includegraphics [width=5cm] {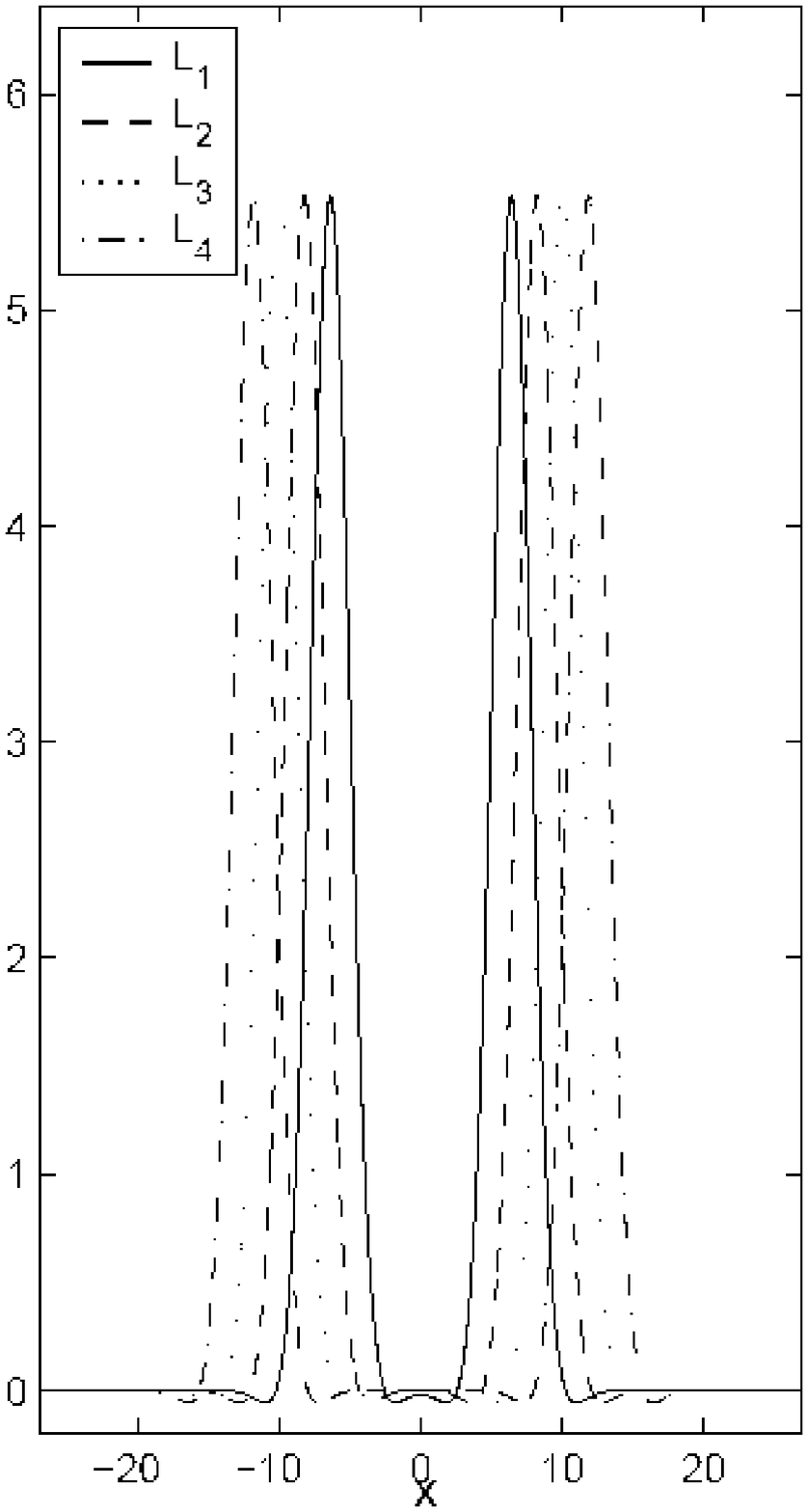}
\end{center}
\caption{Numerical approximation of the first four two-pulse
solutions of the ODE (\ref{ODE}) for $c = 1$ (left) and $c = 4$
(right).}
\end{figure}

We note that comparison of the numerical approximations of two-pulse
solutions $\phi_n(z)$ (obtained with a shooting method) and the
asymptotic approximations from the effective potential
(\ref{effective-potential}) is reported in \cite{CB97}. Their
numerical results (compare Table 1 in \cite{CB97} with our Table 1)
show larger deviations between the two approximations (since their
ODE has a different normalization compared to our (\ref{ODE})) but
the same tendency that the two approximations become identical in
the limit of large separation distances $s_n$ between the two
pulses.

Finally, the three-pulse and multi-pulse solutions of the
fixed-point problem (\ref{ODE}) cannot be approximated numerically
with the use of the Petviashili's method (\ref{Petviashvili}). In
the space of even functions $H^2_{\rm ev}(\mathbb{R})$, the
three-pulse solutions have two finite negative eigenvalues and one
small eigenvalue, while the stabilizing factor of Theorem
\ref{theorem-Petviashvili} and the root finding algorithm of Theorem
\ref{theorem-convergence-two-pulse} can only be useful for one
finite negative eigenvalue and one zero eigenvalue. The additional
finite negative eigenvalue introduces a {\em strong} divergence of
the iterative method (\ref{Petviashvili}) which leads to failure of
numerical approximations for three-pulse solutions. This numerical
problem remains open for further analysis.

\section{Eigenvalues of the stability problem for two-pulse solutions}

We address spectral stability of the two-pulse solution
$\phi_n(z)$ by analyzing the linearized problem
(\ref{stability-problem}), where the operator ${\cal H} :
H^2(\mathbb{R}) \mapsto L^2(\mathbb{R})$ is the Jacobian operator
(\ref{Jacobian}) at $\phi_n(z)$.

By Theorem \ref{theorem-slow-dynamics}(ii), operator ${\cal H}$
has two finite negative eigenvalue, a simple kernel and one small
eigenvalue, which is negative when $W''(L_n) > 0$ and positive
when $W''(L_n) < 0$. Persistence (structural stability) of these
isolated eigenvalues beyond the leading order
(\ref{eigenvalues-H-asymptotic}) is a standard property of
perturbation theory of self-adjoint operators in Hilbert spaces
(see Section IV.3.5 in \cite{Kato}).

By Theorem \ref{theorem-slow-dynamics}(iii), operator $\partial_z
{\cal H}$ has a pair of small eigenvalues, which are purely
imaginary when $W''(L_n) > 0$ and real when $W''(L_n) < 0$. We
first prove that no other eigenvalues may induce instability of
two-pulse solutions (i.e. no other bifurcations of eigenvalues of
$\partial_z {\cal H}$ with ${\rm Re}(\lambda) > 0$ may occur in
$H^2(\mathbb{R})$). We then prove  persistence (structural
stability) of the purely imaginary eigenvalues beyond the leading
order (\ref{eigenvalues-asymptotic}). Combined together, these two
results lead to the theorem on spectral stability of the two-pulse
solution $\phi_n(z)$ which corresponds to $L_n$ with $W''(L_n) >
0$.

\begin{theorem}
\label{theorem-stability} Let $N_{\rm real}$ be the number of real
positive eigenvalues of the linearized problem
(\ref{stability-problem}) in $H^2(\mathbb{R})$, $N_{\rm comp}$ be
the number of complex eigenvalues in the first open quadrant, and
$N_{\rm imag}^-$ be the number of simple positive imaginary
eigenvalues with $({\cal H} v, v) \leq 0$, where $v(x)$ is the
corresponding eigenfunction for $\lambda \in i \mathbb{R}_+$.
Assume that no multiple imaginary eigenvalues exist, the kernel of
${\cal H}$ is simple and $P'(c) > 0$, where $P = \| \phi
\|_{L^2}^2$. Then,
\begin{equation}
\label{closure-relation1} N_{\rm real} + 2 N_{\rm comp} + 2 N_{\rm
imag}^- = n({\cal H}) - 1,
\end{equation}
where $n({\cal H})$ is the number of negative eigenvalues of
${\cal H}$ in $H^2(\mathbb{R})$.
\end{theorem}

\begin{proof}
The statement is equivalent to Theorem 6 in \cite{ChPel} in the
case $({\cal H}^{-1} \phi,\phi) = - (\partial_c \phi,\phi) = -
\frac{1}{2} P'(c) < 0$. The result follows from the Pontryagin
Theorem in Pontryagin space $\Pi_{\kappa}$, where $\kappa =
n({\cal H})$.
\end{proof}

\begin{corollary}
Let $\phi(z) \equiv \Phi(z)$ be a one-pulse solution defined by
Theorem \ref{theorem-one-pulse}. Then, it is a spectrally stable
ground state in the sense that $N_{\rm real} = N_{\rm comp} =
N_{\rm imag}^- = 0$. \label{corollary-one-pulse}
\end{corollary}

\begin{remark}
{\rm It is shown in Lemma 4.12 and Remark 4.14 in \cite{ChPel} that
multiple imaginary eigenvalues may only occur if $({\cal H} v, v) =
0$ such that $n({\cal H}) \geq 2$ is a necessary condition for
existence of multiple eigenvalues (with $P'(c) > 0$). No multiple
imaginary eigenvalues exists for the one-pulse solution $\Phi(z)$.}
\end{remark}

\begin{corollary}
Let $\phi(z) \equiv \phi_n(z)$ be a two-pulse solution defined by
Theorem \ref{theorem-slow-dynamics}. Then,
\begin{itemize}
\item[(i)] the solution corresponding to $L_n$ with $W''(L_n) < 0$
is spectrally unstable in the sense that $N_{\rm real} = 1$ and
$N_{\rm comp} = N_{\rm imag}^- = 0$ for sufficiently small ${\rm
O}(e^{-\kappa_0 L_n})$

\item[(ii)] the solution corresponding to $L_n$ with $W''(L_n) > 0$ satisfies
$N_{\rm real} = 0$ and $N_{\rm comp} + N_{\rm imag}^- = 1$ for
sufficiently small ${\rm O}(e^{-\kappa_0 L_n})$ .
\end{itemize}
\label{corollary-two-pulse}
\end{corollary}

\begin{proof}
It follows from Theorems \ref{theorem-one-pulse} and
\ref{theorem-slow-dynamics} for sufficiently small ${\rm
O}(e^{-\kappa_0 L_n})$ that $P'(c) > 0$, the kernel of ${\cal H}$ is
simple for $W''(L_n) \neq 0$, and the only pair of imaginary
eigenvalues with $({\cal H} v, v) < 0$ in the case $W_n''(L_n)
> 0$ is simple. Therefore, assumptions of Theorem \ref{theorem-stability} are
satisfied for the two-pulse solutions $\phi_n(z)$ with $W''(L_n)
\neq 0$. By the count of Theorem \ref{theorem-slow-dynamics}(ii),
$n({\cal H}) = 3$ for $W''(L_n) > 0$ and $n({\cal H}) = 2$ for
$W''(L_n) < 0$. Furthermore, persistence (structural stability) of
simple real eigenvalues of the operator $\partial_z {\cal H}$ in
$H^2(\mathbb{R})$ follows from the perturbation theory of isolated
eigenvalues of non-self-adjoint operators (see Section VIII.2.3 in
\cite{Kato}).
\end{proof}

There exists one uncertainty in Corollary
\ref{corollary-two-pulse}(ii) since it is not clear if the
eigenvalue of negative Krein signature in Theorem
\ref{theorem-slow-dynamics}(iii) remains imaginary in $N_{\rm
imag}^-$ or bifurcates to a complex eigenvalue in $N_{\rm comp}$.
This question is important for spectral stability of the
corresponding two-pulse solutions since the former case implies
stability while the latter case implies instability of solutions.
We will remove the uncertainty and prove that $N_{\rm imag}^- = 1$
and $N_{\rm comp} = 0$ for sufficiently small ${\rm
O}(e^{-\kappa_0 L})$. To do so, we rewrite the linearized problem
(\ref{stability-problem}) in the exponentially weighted space
\cite{PW94}:
\begin{equation}
\label{exponential-weight} H^2_{\alpha} = \left\{ v \in H^2_{\rm
loc}(\mathbb{R}) : \;\; e^{\alpha z} v(z) \in H^2(\mathbb{R})
\right\}.
\end{equation}
The linearized operator $\partial_z {\cal H}$ transforms to the form
\begin{equation}
\label{operator-exponential} {\cal L}_{\alpha} = (\partial_z -
\alpha) \left( c - (\partial_z - \alpha)^2 + (\partial_z - \alpha)^4
- 2 \phi(z) \right),
\end{equation}
which acts on the eigenfunction $v_{\alpha}(z) = e^{\alpha z} v(z)
\in H^2(\mathbb{R})$. The absolute continuous part of the spectrum
of ${\cal L}_{\alpha}$ is located at $\lambda =
\lambda_{\alpha}(k)$, where
\begin{equation}
\lambda_{\alpha}(k) = (i k - \alpha) (c - (ik - \alpha)^2 + (i k -
\alpha)^4), \qquad k \in \mathbb{R}.
\end{equation}
A simple analysis shows that
\begin{eqnarray*}
\frac{d}{dk} {\rm Re}(\lambda_{\alpha}(k)) & = & -2 \alpha k (10 k^2
- 10 \alpha^2 + 3), \\
\frac{d}{dk} {\rm Im}(\lambda_{\alpha}(k)) & = & c - 3 \alpha^2 + 5
\alpha^4 + 3 k^2 (1 - 10 \alpha^2) + 5 k^4.
\end{eqnarray*}
The following lemma gives a precise location of the dispersion
relation $\lambda = \lambda_{\alpha}(k)$ on $\lambda \in
\mathbb{C}$.

\begin{lemma}
The dispersion relation $\lambda = \lambda_{\alpha}(k)$ is a
simply-connected curve located in the left half-plane of $\lambda
\in \mathbb{C}$ if
\begin{equation}
\label{alpha-constraint} 0 < \alpha < \frac{1}{\sqrt{10}}, \qquad
c > \frac{1}{4}.
\end{equation}
\end{lemma}

\begin{proof}
The mapping $k \mapsto {\rm Im}(\lambda_{\alpha})$ is one-to-one
provided that $c - 3 \alpha^2 + 5 \alpha^4 > 0$ and $1 - 10 \alpha^2
> 0$. Since $c - 3 \alpha^2 + 5 \alpha^4$ reaches the minimum value
on $\alpha \in \left[ 0, \frac{1}{\sqrt{10}}\right]$ at the right
end $\alpha = \frac{1}{\sqrt{10}}$ and the minimum value is positive
if $c > \frac{1}{4}$, the first inequality is satisfied under
(\ref{alpha-constraint}). The second inequality is obviously
satisfied if $|\alpha | < \frac{1}{\sqrt{10}}$. The mapping $k
\mapsto {\rm Re}(\lambda_{\alpha})$ has a single extremal point at
$k = 0$ provided $3 - 10 \alpha^2 > 0$, which is satisfied if
$|\alpha |< \frac{1}{\sqrt{10}}$. The extremal point is the point of
maximum and the entire curve is located in the left half-plane of
$\lambda \in \mathbb{C}$ if $0 < \alpha < \frac{1}{\sqrt{10}}$.
\end{proof}

The following two lemmas postulate properties of eigenfunctions
corresponding to embedded eigenvalues of negative Krein signature.

\begin{lemma}
\label{lemma-persistence} Let $v_0(z)$ be an eigenfunction of
$\partial_z {\cal H}$ for a simple eigenvalue $\lambda_0 \in i
\mathbb{R}_+$ in $H^2(\mathbb{R})$. Then, $\lambda_0 \in i
\mathbb{R}_+$ is also an eigenvalue in $H^2_{\alpha}(\mathbb{R})$
for sufficiently small $\alpha$.
\end{lemma}

\begin{proof}
Let $k = k_0 \in \mathbb{R}$ be the unique real root of the
dispersion relation $\lambda_0(k) = \lambda_0$ (with $\alpha = 0$)
for a given eigenvalue $\lambda_0 \in i \mathbb{R}_+$. The other
four roots $k = k_{1,2,3,4}$ for a given $\lambda_0 \in i
\mathbb{R}_+$ are complex with $|{\rm Re}(k_j)| \geq \kappa_0 > 0$.
By the Stable and Unstable Manifolds Theorem in linearized ODEs
\cite{CodLev}, the decaying eigenfunction $v_0(z) \in
H^2(\mathbb{R})$ is exponentially decaying with the decay rate
greater than $\kappa_0 > 0$ and it does not include the bounded term
$e^{i k_0 z}$ as $z \to \pm \infty$.  By construction,
$v_{\alpha}(z) = e^{\alpha z} v_0(z)$ is also exponentially decaying
as $z \to \pm \infty$ for sufficiently small $|\alpha| < \kappa_0$.
Since $v_0 \in H^2(\mathbb{R})$ and due to the exponential decay of
$v_{\alpha}(z)$ as $|z| \to \infty$, we have $v_{\alpha} \in
H^2(\mathbb{R})$ for any small $\alpha$.
\end{proof}

\begin{lemma}
\label{lemma-symmetry} Let $v_0(z) \in H^2(\mathbb{R})$ be an
eigenfunction of $\partial_z {\cal H}$ for a simple eigenvalue
$\lambda_0 \in i \mathbb{R}_+$ with $({\cal H} v_0, v_0) < 0$.
Then, there exists $w_0 \in H^2(\mathbb{R})$, such that $v_0 =
w_0'(x)$ and $w_0(z)$ is an eigenfunction of ${\cal H}
\partial_z$ for the same eigenvalue $\lambda_0$. Moreover,
$(w_0,v_0) \in i \mathbb{R}_+$.
\end{lemma}

\begin{proof}
Since ${\cal H} : H^2(\mathbb{R}) \mapsto L^2(\mathbb{R})$, the
eigenfunction $v_0(z)$ of the eigenvalue problem $\partial_z {\cal
H} v_0 = \lambda_0 v_0$ for any $\lambda_0 \neq 0$ must satisfy the
constraint $\int_{\mathbb{R}} v_0(z) dz = 0$. Let $v_0 = w_0'(z)$.
Since $v_0(z)$ decays exponentially as $|z| \to \infty$ and $(1,v_0)
= 0$, then $w_0(z)$ decays exponentially as $|z| \to \infty$, so
that $w_0 \in H^2(\mathbb{R})$. By construction, ${\cal H}
\partial_z w_0 = {\cal H} v_0 = \lambda_0 \int v_0(z) dz = \lambda
w_0$. The values of $(w_0,v_0)$ are purely imaginary as
$$
\overline{(w_0,v_0)} = \int_{\mathbb{R}} \bar{w}_0 v_0 dz =
\int_{\mathbb{R}} \bar{w}_0 \partial_z w_0 dz = - \int_{\mathbb{R}}
w_0 \partial_z \bar{w}_0 dz = -\int_{\mathbb{R}} w_0 \bar{v}_0 dz =
- (w_0,v_0).
$$
Since ${\cal H} v_0 = \lambda_0 w_0$ with $\lambda_0 \in i
\mathbb{R}_+$ and $({\cal H} v_0, v_0) < 0$, we have $(w_0,v_0) =
\lambda_0^{-1} ({\cal H} v_0,v_0) \in i \mathbb{R}_+$.
\end{proof}

The following theorem states that the embedded eigenvalues of
negative Krein signature are structurally stable in the linearized
problem (\ref{stability-problem}).

\begin{theorem}
Let $\lambda_0 \in \mathbb{R}_+$ be a simple eigenvalue of
$\partial_z {\cal H}$ with the eigenfunction $v_0 \in
H^2(\mathbb{R})$ such that $({\cal H} v_0, v_0) < 0$. Then, it is
structurally stable to parameter continuations, e.g. for any $V
\in L^{\infty}(\mathbb{R})$ and sufficiently small $\delta$, there
exists an eigenvalue $\lambda_{\delta} \in i \mathbb{R}_+$ of
$\partial_z \left( {\cal H} + \delta V(z) \right)$ in
$H^2(\mathbb{R})$, such that $|\lambda_{\delta} - \lambda_0| \leq
C \delta$ for some $C > 0$. \label{theorem-persistence}
\end{theorem}

\begin{proof}
By Lemma \ref{lemma-persistence}, $\lambda_0$ is also an eigenvalue
of ${\cal L}_{\alpha}$ in $H^2(\mathbb{R})$ for sufficiently small
$\alpha$. Let $\alpha$ be fixed in the bound
(\ref{alpha-constraint}). There exists a small neighborhood of
$\lambda_0$, which is isolated from the absolute continuous part of
the spectrum of ${\cal L}_{\alpha}$. By the perturbation theory of
isolated eigenvalues of non-self-adjoint operators (see Section
VIII.2.3 in \cite{Kato}), there exists a simple eigenvalue
$\lambda_{\delta}$ of $\partial_z ({\cal H} + \delta V(z))$ in
$H^2_{\alpha}(\mathbb{R})$ for the same value of $\alpha$ and
sufficiently small $\delta$ in a local neighborhood of $\lambda_0$,
such that $|\lambda_{\delta} - \lambda_0| \leq C \delta$ for some $C
> 0$.

It remains to show that the simple eigenvalue $\lambda_{\delta}$
is purely imaginary for the same value of $\alpha > 0$. Denote the
eigenfunction of $\partial_z ({\cal H} + \delta V(z))$ in
$H^2_{\alpha}(\mathbb{R})$ for the eigenvalue $\lambda_{\delta}$
by $v_{\delta}(z)$, such that $e^{\alpha z} v_{\delta} \in
H^2(\mathbb{R})$. If $v_{\delta} \notin H^2(\mathbb{R})$, then the
count of eigenvalues (\ref{closure-relation1}) in
$H^2(\mathbb{R})$ is discontinuous at $\delta = 0$: the eigenvalue
$\lambda_0$ in the number $N_{\rm imag}^-$ at $\delta = 0$
disappears from the count for $\delta \neq 0$. If $v_{\delta} \in
H^2(\mathbb{R})$, then $(1,v_{\delta}) = 0$ and since
$v_{\delta}(z)$ is exponentially decaying as $|z| \to \infty$,
there exists $w_{\delta}(z) \in H^2(\mathbb{R})$ such that
$v_{\delta} = w_{\delta}'(z)$. The 2-form
$(w_{\delta},v_{\delta})$ is invariant with respect to the weight
$\alpha$ since if $e^{\alpha z} v_{\delta}(z)$ is an eigenfunction
of $\partial_z ({\cal H} + \delta V(z))$ in $H^2(\mathbb{R})$ for
the eigenvalue $\lambda_{\delta}$ (i.e. $v_{\delta} \in
H^2_{\alpha}(\mathbb{R})$), then $e^{-\alpha z} w_{\delta}(z)$ is
an eigenfunction of $({\cal H} + \delta V(z))
\partial_z$ in $H^2(\mathbb{R})$ for the same eigenvalue
$\lambda_{\delta}$ (i.e. $w_{\delta} \in
H^2_{-\alpha}(\mathbb{R})$). Computing $(w_{\delta},v_{\delta})$ at
$\alpha = 0$, we have
$$
\lambda_{\delta} (w_{\delta},v_{\delta}) = ({\cal H}
v_{\delta},v_{\delta}) \in \mathbb{R},
$$
which indicates that $(w_{\delta},v_{\delta}) \equiv 0$ if ${\rm
Re}(\lambda_{\delta}) \neq 0$ and ${\rm Im}(\lambda_{\delta}) >
0$. Since $(w_0,v_0) \neq 0$ by Lemma \ref{lemma-symmetry} and
$(w_{\delta},v_{\delta})$ is continuous in $\delta$, the case
${\rm Re}(\lambda_{\delta}) \neq 0$ is impossible. Therefore,
${\rm Re}(\lambda_{\delta}) = 0$ is preserved for small non-zero
$\delta$.
\end{proof}

\begin{corollary}
Let $\phi(z) \equiv \phi_n(z)$ be a two-pulse solution defined by
Theorem \ref{theorem-slow-dynamics} that corresponds to $L_n$ with
$W''(L_n) > 0$. Then, it is spectrally stable in the sense that
$N_{\rm real} = N_{\rm comp} = 0$ and $N_{\rm imag}^- = 1$ for
sufficiently small ${\rm O}(e^{-\kappa_0 L_n})$.
\label{corollary-two-pulse-definite}
\end{corollary}

\begin{remark}
{\rm Using perturbation theory in $H^2_{\alpha}(\mathbb{R})$ for a
fixed value $\alpha > 0$, one cannot a priori exclude the shift of
eigenvalue $\lambda_0$ to $\lambda_{\delta}$ with ${\rm
Re}(\lambda_{\delta}) > 0$. Even if $v_0(z)$ for $\lambda_0$
contains no term $e^{i k_0 z}$ as $z \to -\infty$ (see Lemma
\ref{lemma-persistence}), the eigenfunction $v_{\delta}(z)$ for
$\lambda_{\delta}$ may contain the term $e^{i k_{\delta} z}$ as $z
\to -\infty$ with ${\rm Im}(k_{\delta}) < 0$ and $\lim_{\delta \to
0} k_{\delta} = k_0 \in \mathbb{R}$. However, when Theorem
\ref{theorem-persistence} holds (that is under the assumptions
that $v_0 \in H^2(\mathbb{R})$ and $({\cal H} v_0,v_0) < 0$), the
eigenvalue $\lambda_{\delta}$ remains on $i \mathbb{R}$ and the
eigenfunction $v_{\delta}(z)$ must have no term $e^{i k_{\delta}
z}$ with $k_{\delta} \in \mathbb{R}$ as $z \to -\infty$ for any
sufficiently small $\delta$. The hypothetical bifurcation above
can however occur if $v_0 \notin H^2(\mathbb{R})$ but $v_0 \in
H^2_{\alpha}(\mathbb{R})$ with $\alpha > 0$. We do not know any
example of such bifurcation in the case $v_0 \notin
H^2(\mathbb{R})$ but $({\cal H} v_0,v_0) \geq 0$. }
\label{remark-bifurcation}
\end{remark}

\begin{remark}
{\rm When the potential is symmetric (i.e. $\phi(-z) = \phi(z)$),
there exists a remarkable symmetry of eigenvalues and
eigenfunctions of the stability problem $\partial_z {\cal H} v =
\lambda v$: if $v(z)$ is an eigenfunction for $\lambda$, then
$\overline{v(-z)}$ is the eigenfunction for $-\bar{\lambda}$. If
$\lambda_0 \in i \mathbb{R}$ is a simple eigenvalue and $v_0 \in
H^2_{\alpha}(\mathbb{R})$ with $\alpha \geq 0$, the above symmetry
shows that $v_0 \in H^2_{-\alpha}(\mathbb{R})$ with $-\alpha \leq
0$. If ${\rm Re}(\lambda_{\delta}) > 0$ and $v_{\delta} \in
H^2_{\alpha}(\mathbb{R})$, then $-\overline{v_{\delta}(-z)}$ is an
eigenfunction of the same operator in $H^2_{-\alpha}(\mathbb{R})$
with $-\alpha \leq 0$ for eigenvalue ${\rm
Re}(-\overline{\lambda_{\delta}}) = -{\rm Re}(\lambda_{\delta})$
and ${\rm Im}(-\overline{\lambda_{\delta}}) = {\rm
Im}(\lambda_{\delta})$. Thus, the hypothetical bifurcation in
Remark \ref{remark-bifurcation} implies that the embedded
eigenvalue $\lambda_0 \in i \mathbb{R}$ may split into two
isolated eigenvalues $\lambda_{\delta}$ and
$-\overline{\lambda_{\delta}}$ as $\delta \neq 0$. Theorem
\ref{theorem-persistence} shows that such splitting is impossible
in $H^2(\mathbb{R})$ if $({\cal H} v_0,v_0) < 0$. }
\end{remark}

We confirm results of Corollaries \ref{corollary-two-pulse} and
\ref{corollary-two-pulse-definite} with numerical computations of
eigenvalues in the linearized problem (\ref{stability-problem}).
Throughout computations, we use the values $\alpha = 0.04$ and $c
= 1$, which satisfy the constraint (\ref{alpha-constraint}). The
spectra of the operators ${\cal H}$ in $H^2(\mathbb{R})$ and
$\partial_z {\cal H}$ in $H^2_{\alpha}(\mathbb{R})$ are computed
by using the Fourier spectral method. This method is an obvious
choice since the solution $\phi(z)$ is obtained by using the
spectral approximations in the iterative scheme
(\ref{Petviashvili})--(\ref{factor}). As in the previous section,
we use numerical parameters $d = 100$, $h = 0.01$ and $\varepsilon
= 10^{-15}$ for the Petviashvili's method
(\ref{Petviashvili})--(\ref{factor}).

Eigenvalues of the discretized versions of the operators ${\cal
H}$ and ${\cal L}_{\alpha}$ are obtained with the MATLAB
eigenvalue solver \verb"eig". The spectra are shown on Figure 6
for the two-pulse solution $\phi_1(z)$ and on Figure 7 for the
two-pulse solution $\phi_2(z)$. The inserts show zoomed
eigenvalues around the origin and the dotted line connects
eigenvalues of the discretized operators that belong to the
absolutely continuous part of the spectra. Figures 6 and 7 clearly
illustrate that the small eigenvalue of ${\cal H}$ is negative for
$\phi_1(z)$ and positive for $\phi_2(z)$, while the pair of small
eigenvalues of ${\cal L}_{\alpha}$ is purely imaginary for
$\phi_1(z)$ and purely real for $\phi_2(z)$. This result is in
agreement with Corollaries \ref{corollary-two-pulse} and
\ref{corollary-two-pulse-definite}. We have observed the same
alternation of small eigenvalues for two-pulse solutions
$\phi_3(z)$ and $\phi_4(z)$, as well as for other values of
parameters $c$ and $\alpha$.

The numerical discretization based on the Fourier spectral method
shifts eigenvalues of the operators ${\cal H}$ and ${\cal
L}_{\alpha}$. In order to measure the numerical error introduced
by the discretization, we compute the numerical value for the
"zero" eigenvalue corresponding to the simple kernel of ${\cal H}$
and the double zero eigenvalue of ${\cal L}_{\alpha}$. Table II
shows numerical values for the "zero" and small eigenvalues for
two-pulse solutions $\phi_n(z)$ with $n = 1,2,3,4$. It is obvious
from the numerical data that the small eigenvalues are still
distinguished (several orders higher) than the numerical
approximations for zero eigenvalues for $n = 1,2,3$ but they
become comparable for higher-order two-pulse solutions $n \geq 4$.
This behavior is understood from Theorem
\ref{theorem-slow-dynamics} since the small eigenvalues becomes
exponentially small for larger values of $s$ (larger $n$) in the
two-pulse solution (\ref{decomposition}) and the exponentially
small contribution is negligible compared to the numerical error
of discretization.

\begin{center}
\begin{tabular}{|c|l|l|l|l|}
    \hline
& $\phi_1(z)$ & $\phi_2(z)$ & $\phi_3(z)$ & $\phi_4(z)$  \\
    \hline
"Zero" eigenvalue of ${\cal H}$  & $1.2156 \cdot 10^{-9}$   &
$2.6678 \cdot 10^{-9}$
& $1.4742 \cdot 10^{-9}$  & $1.8941 \cdot 10^{-9}$ \\
    \hline
Small eigenvalue of ${\cal H}$ & $1.7845 \cdot 10^{-2}$   & $7.6638 \cdot 10^{-5}$
& $3.3342 \cdot 10^{-7}$  & $2.9212 \cdot 10^{-9}$ \\
    \hline
"Zero" eigenvalues of ${\cal L}_{\alpha}$ & $0.3653 \cdot 10^{-5}$ &
$0.5321 \cdot 10^{-5}$
& $0.7832 \cdot 10^{-5}$  & $1.2374 \cdot 10^{-5}$ \\
    \hline
${\rm Re}$ of small eigenvalues of ${\cal L}_{\alpha}$ & $4.5291
\cdot 10^{-6}$ & $3.2845 \cdot 10^{-3}$ & $6.3256 \cdot 10^{-5}$  & $1.6523 \cdot 10^{-5}$ \\
    \hline
${\rm Im}$ of small eigenvalues of ${\cal L}_{\alpha}$ & $0.5021
\cdot 10^{-1}$ & $1.1523 \cdot 10^{-8}$ & $2.1672 \cdot 10^{-4}$  & $5.4443 \cdot 10^{-6}$ \\
    \hline
\end{tabular}
\end{center}

{\bf Table II:} Numerical approximations of the zero and small
eigenvalues of operators ${\cal H}$ and ${\cal L}_{\alpha}$ for
the first four two-pulse solutions with $c = 1$, $\alpha = 0.04$,
$d = 100$, $h = 0.01$ and $\varepsilon = 10^{-15}$. The absolute
values of real and imaginary eigenvalues are shown.

\begin{figure}
\begin{center}
\includegraphics [width=9cm]{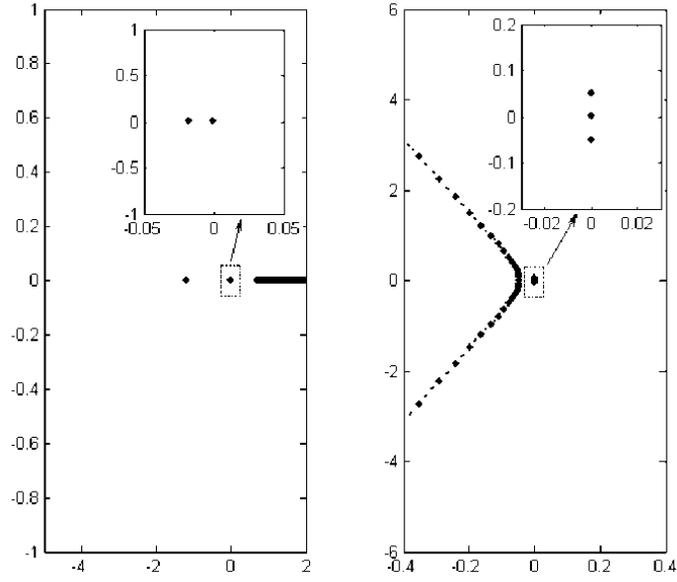}
\end{center}
\caption{Numerical approximations of the spectra of operators
${\cal H}$ and ${\cal L}_{\alpha}$ for the two-pulse solution
$\phi_1(z)$ with $c = 1$ and $\alpha = 0.04$. The insert shows
zoom of small eigenvalues and the dotted curve connects
eigenvalues of the continuous spectrum of ${\cal L}_{\alpha}$.}
\end{figure}

\begin{figure}
\begin{center}
\includegraphics [width=9cm]{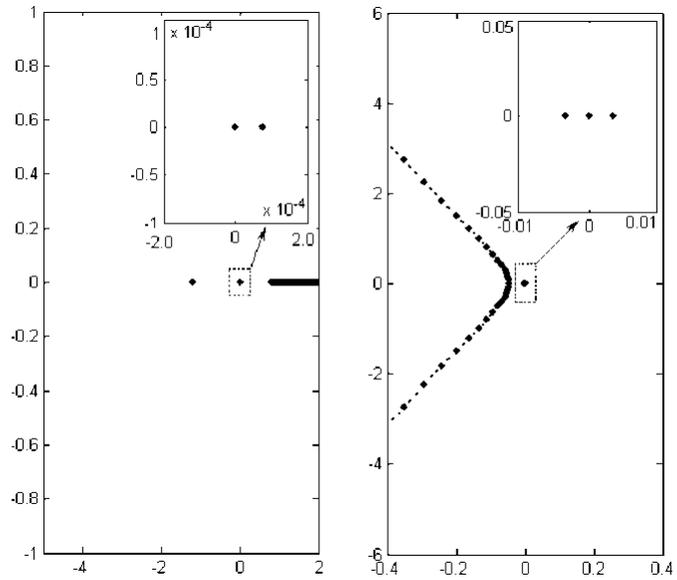}
\end{center}
\caption{The same as Figure 6 but for the two-pulse solution
$\phi_2(z)$.}
\end{figure}

We have confirmed numerically the analytical predictions that all
two-pulse solutions corresponding to the points $L_n$ with $W''(L_n)
< 0$ (which are maxima of the effective interaction potential) are
unstable with a simple real positive eigenvalue, while all two-pulse
solutions corresponding to the points $L_n$ with $W''(L_n)
> 0$ (which are minima of the effective interaction potential) are
spectrally stable. The stable two-pulse solutions are not however
ground states since the corresponding linearized problem has a
pair of purely imaginary eigenvalues of negative Krein signature.

\section{Nonlinear dynamics of two-pulse solution}

The Newton's particle law (\ref{Newton-law}) is a useful
qualitative tool to understand the main results of our article.
Existence of an infinite countable sequence of two-pulse solution
$\{\phi_n(z)\}_{n \in \mathbb{Z}}$ is related to existence of
extremal points $\{ L_n \}_{n \in \mathbb{Z}}$ of the effective
potential function $W(L)$, while alternation of stability and
instability of the two-pulse solutions is related to the
alternation of minima and maxima points of $W(L)$. It is natural
to ask if the Newton's law (\ref{Newton-law}) extends beyond the
existence and spectral stability analysis of two-pulse solutions
in the fifth-order KdV equation (\ref{kdv}). In particular, one
can ask if the purely imaginary (embedded) eigenvalues of the
linearized problem (\ref{stability-problem}) lead to nonlinear
asymptotic stability of two-pulse solutions or at least to their
nonlinear stability in the sense of Lyapunov. From a more
technical point of view, one can ask whether the Newton's law
(\ref{Newton-law}) serves as the center manifold reduction for
slow nonlinear dynamics of two-pulse solutions in the PDE
(\ref{kdv}) and whether solutions of the full problem are
topologically equivalent to solutions of the Newton's law. While
we do not attempt to develop mathematical analysis of these
questions, we illustrate nonlinear dynamics of two-pulse solutions
with explicit numerical simulations.

The numerical pseudo-spectral method for solutions of the
fifth-order KdV equation (\ref{kdv}) is described in details in
\cite{MV}. The main idea of this method is to compute analytically
the linear part of the PDE (\ref{kdv}) by sing the Fourier
transform and to compute numerically its nonlinear part by using
an ODE solver. Let $\hat{u}(k,t)$ denote the Fourier transform of
$u(x,t)$ and rewrite the PDE (\ref{kdv}) in the Fourier domain:
\begin{equation}
\label{kdv-Fourier} \hat{u}_t  = i(k^3 + k^5)\hat{u} - ik
\widehat{u^2}.
\end{equation}
In order to compute $\widehat{u^2}(k,t)$ we evaluate $u^2(x,t)$ on
$x \in \mathbb{R}$ and apply the discrete Fourier transform.
Substitution $\hat{u} = s(k,t) e^{i(k^3 + k^5)t}$ transforms the
evolution equation (\ref{kdv-Fourier}) to the form:
\begin{equation}
\label{ODE-Fourier} s_t = -i k e^{-i(k^3+k^5)t}\widehat{u^2}(k,t).
\end{equation}
The fourth-order Runge-Kutta method is used to integrate the
evolution equation (\ref{ODE-Fourier}) in time with time step
$\triangle t$. To avoid large variations of the exponent for large
values of $k$ and $t$, the substitution above is updated after $m$
time steps as follows:
\begin{equation}
\hat{u} = s_m(k,t) e^{i(k^3 + k^5)(t - m \triangle t)}, \qquad m
\triangle t \leq t \leq (m+1) \triangle t.
\end{equation}
The greatest advantage of this numerical method is that no
stability restriction arising from the linear part of
(\ref{kdv-Fourier}) is posed on the timestep of numerical
integration. On contrast, the standard explicit method for the
fifth-order KdV equation (\ref{kdv}) has a serious limitation on
the timestep of the numerical integration since the fifth-order
derivative term brings stiffness to the evolution problem. The
small timestep would be an obstacle for the long time integration
of the evolution problem due to accumulation of computational
errors.

Numerical simulations of the PDE (\ref{kdv-Fourier}) are started
with the initial condition:
\begin{equation}
\label{initial-condition-PDE} u(x,0) = \Phi(x-s) + \Phi(x+s),
\end{equation}
where $\Phi(x)$ is the one-pulse solution and $2s$ is the initial
separation between the two pulses. The one-pulse solution
$\Phi(x)$ is constructed with the iteration method
(\ref{Petviashvili})--(\ref{factor}) for $c = 4$. The numerical
factors of the spectral approximation are $L = 100$,  $N =
2^{12}$, $\varepsilon = 10^{-15}$, while the timestep is set to
$\triangle t = 10^{-4}$.

Figure 8 shows six individual simulations of the initial-value
problem (\ref{kdv-Fourier}) and (\ref{initial-condition-PDE}) with
$s = 2.3$, $s = 2.8$, $s = 3.6$, $s = 4.2$, $s = 4.5$ and $s =
4.7$. Figure 9 brings these six individual simulations on the
effective phase plane $(L,\dot{L})$ computed from the distance
$L(t)$ between two local maxima (humps) of the two-pulse
solutions.

When the initial distance $(s = 2.3)$ is taken far to the left
from the stable equilibrium point (which corresponds to the
two-pulse solution $\phi_1(x)$), the two pulses repel and diverge
from each other (trajectory 1). When the initial distance $(s =
2.8)$ is taken close to the left from the stable equilibrium
point, we observe small-amplitude oscillations of two pulses
relative to each other (trajectory 2). When the initial distances
$(s = 3.6)$ and $(s = 4.2)$ are taken to the right from the stable
equilibrium point, we continue observing stable oscillations of
larger amplitudes and larger period (trajectories 3 and 4). The
oscillations are destroyed when the initial distances are taken
close to the unstable equilibrium point (which corresponds to the
two-pulse solution $\phi_2(x)$) from either left $(s = 4.5)$ or
right $(s = 4.7)$. In either case, the two pulses repel and
diverge from each other (trajectories 5 and 6). Ripples in the
pictures are due to radiation effect and the numerical integration
does not make sense after $t \approx 500$ because the ripples
reach the left end of the computational interval and appear from
the right end due to periodic boundary conditions.

The numerical simulations of the full PDE problem (\ref{kdv})
indicate the validity of the Newton's particle law
(\ref{Newton-law}). Due to the energy conservation, all
equilibrium points in the Newton's law are either centers or
saddle points and the center points are surrounded by closed
periodic orbits in the interior of homoclinic loops from the
stable and unstable manifolds of the saddle points. Trajectories
2,3, and 4 are taken inside the homoclinic orbit from the saddle
point corresponding to $\phi_2(x)$ and these trajectories
represent periodic oscillations of two-pulse solutions near the
center point corresponding to $\phi_1(x)$. Trajectories 1 and 6
are taken outside the homoclinic orbit and correspond to unbounded
dynamics of two-pulse solutions. The only exception from the
Newton's law (\ref{Newton-law}) is trajectory 5, which is supposed
to occur inside the homoclinic loop but turns out to occur outside
the loop. This discrepancy can be explained by the fact that the
Newton's law (\ref{Newton-law}) does not {\em exactly} represent
the dynamics of the PDE (\ref{kdv-Fourier}) generated by the
initial condition (\ref{initial-condition-PDE}) but it corresponds
to an {\em asymptotic} solution after the full solution is
projected into the discrete and continuous parts and the
projection equations are truncated (see details in \cite{Sigal} in
the context of the NLS equations).

\begin{figure}
\begin{center}
\includegraphics[width = 7cm]{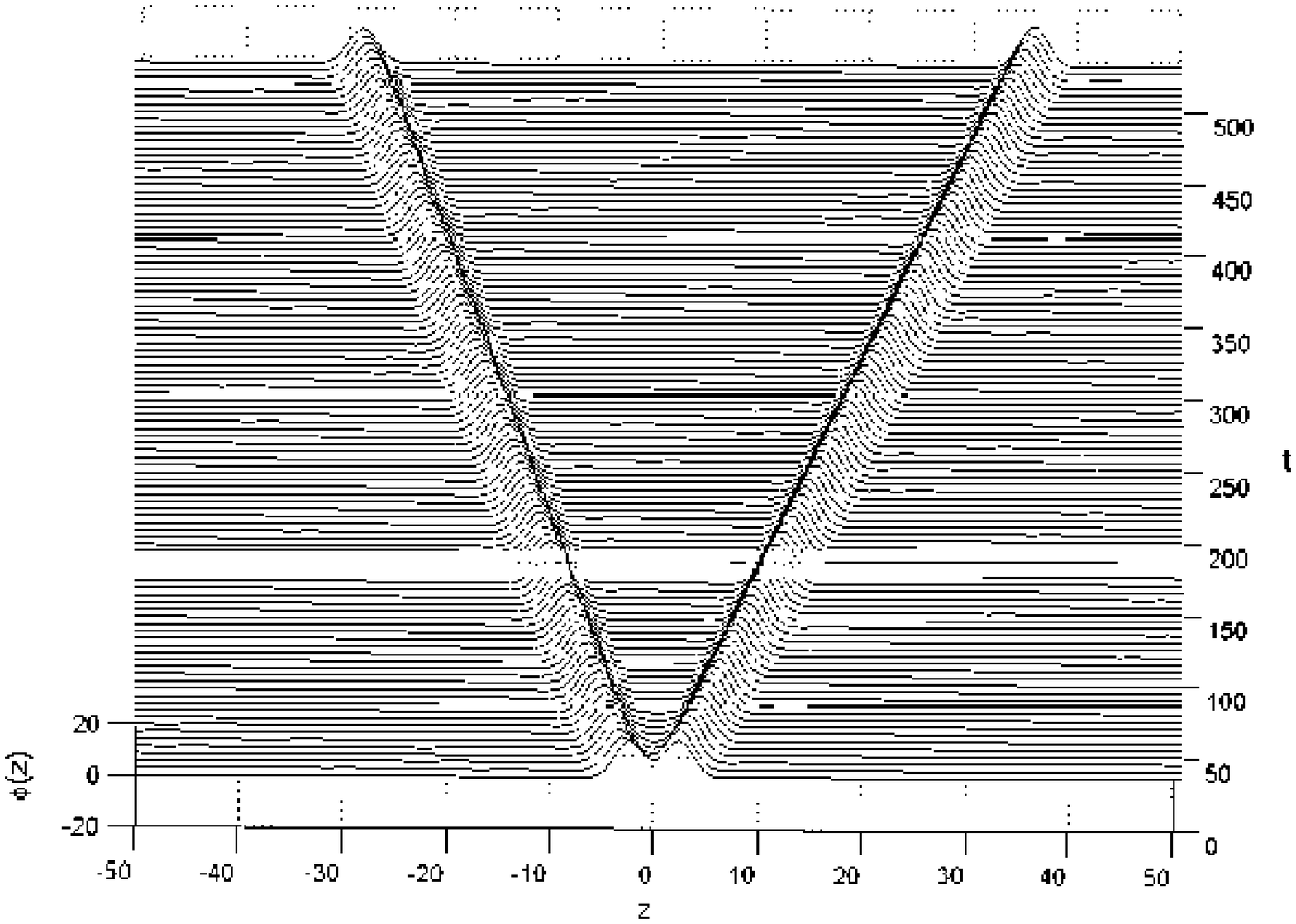}
\includegraphics[width = 7cm]{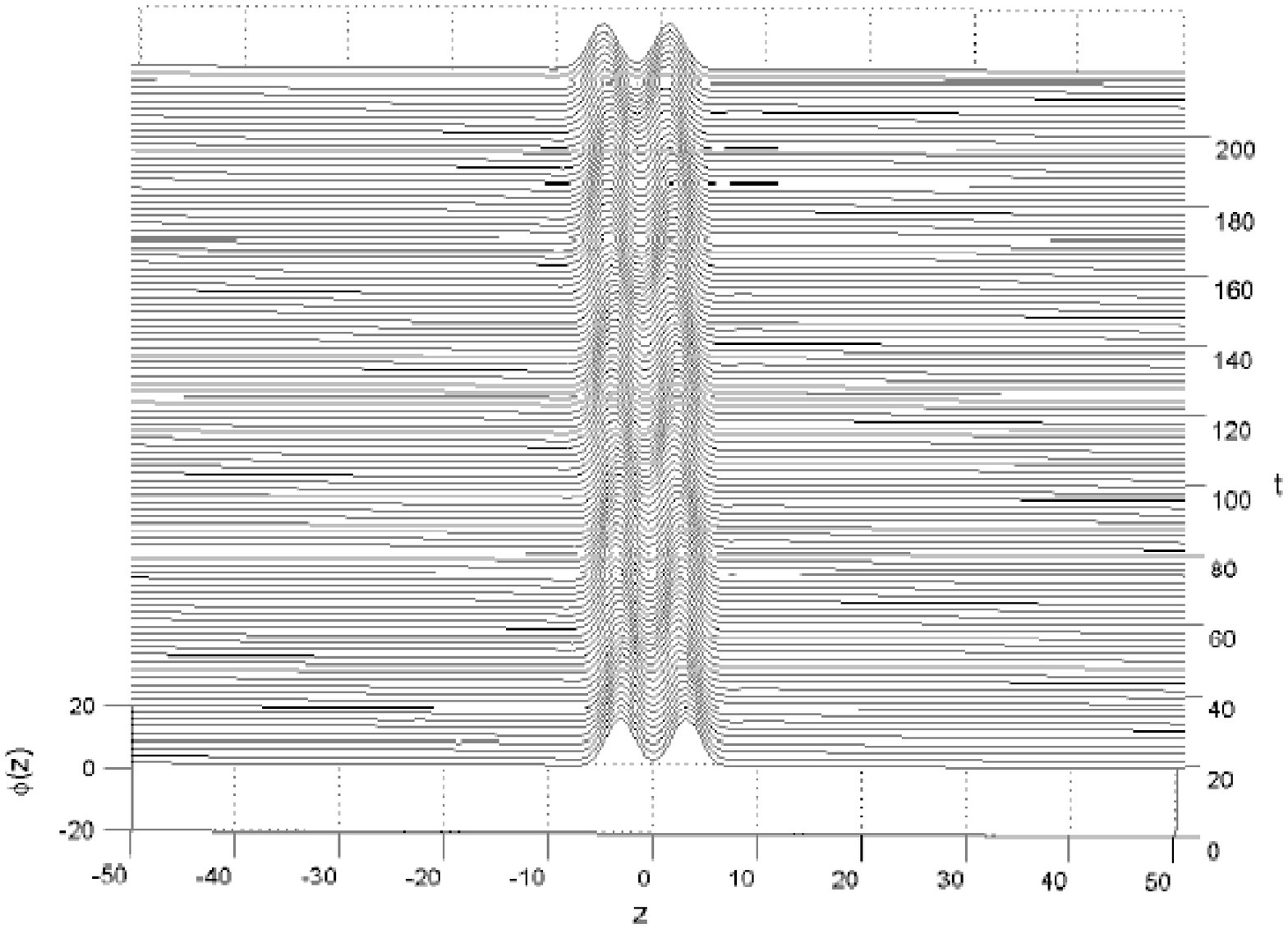}
\includegraphics[width = 7cm]{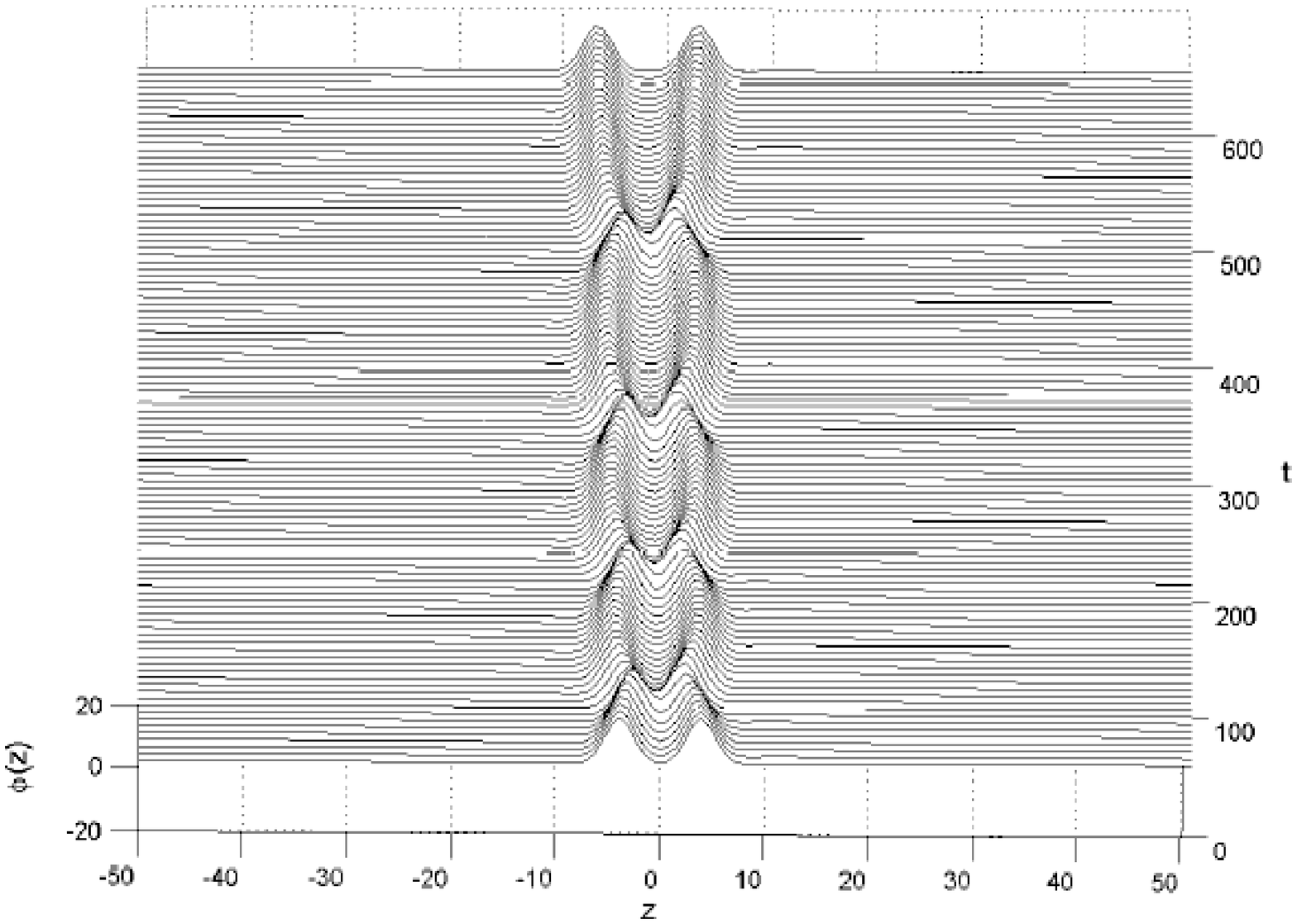}
\includegraphics[width = 7cm]{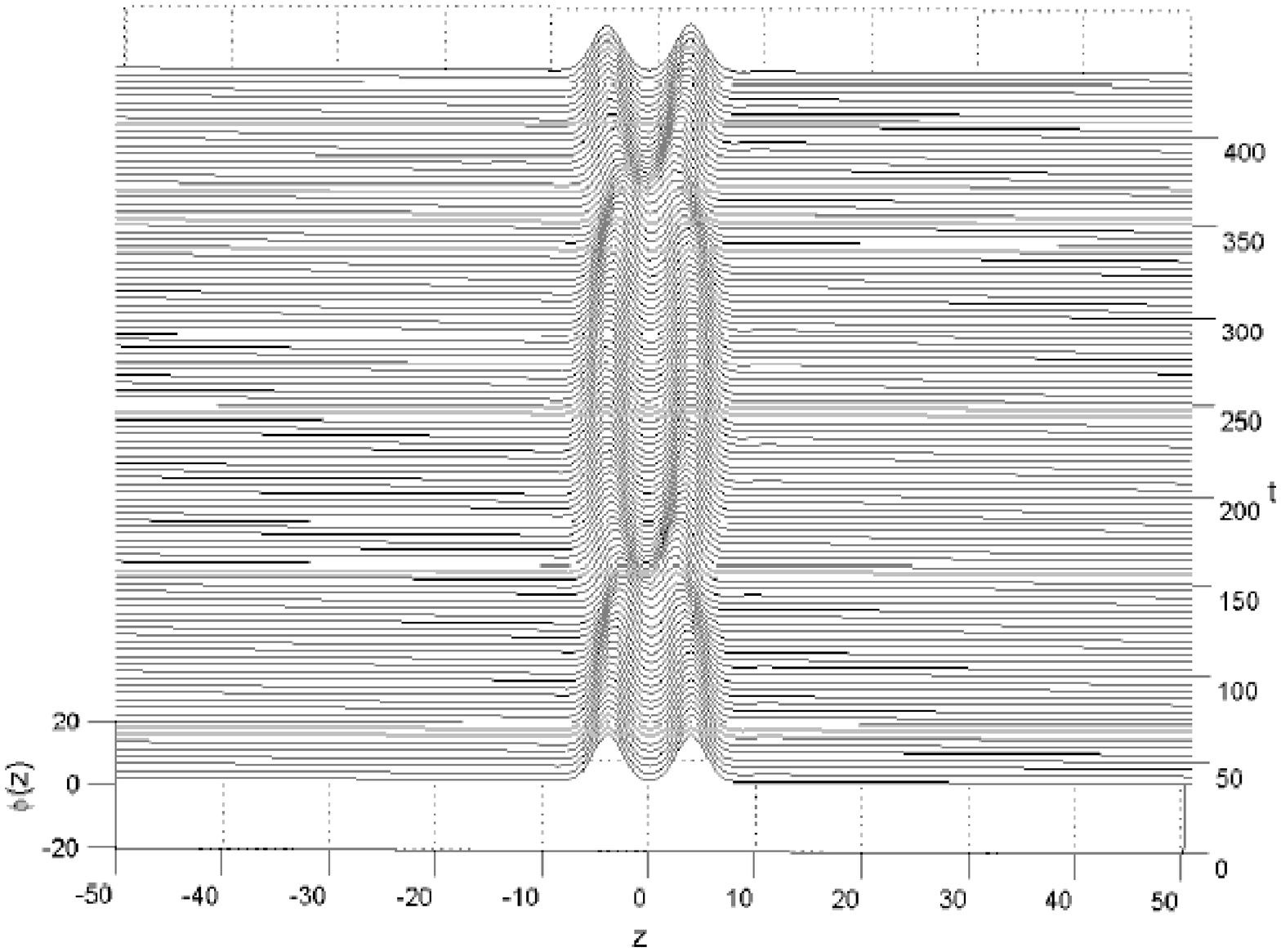}
\includegraphics[width = 7cm]{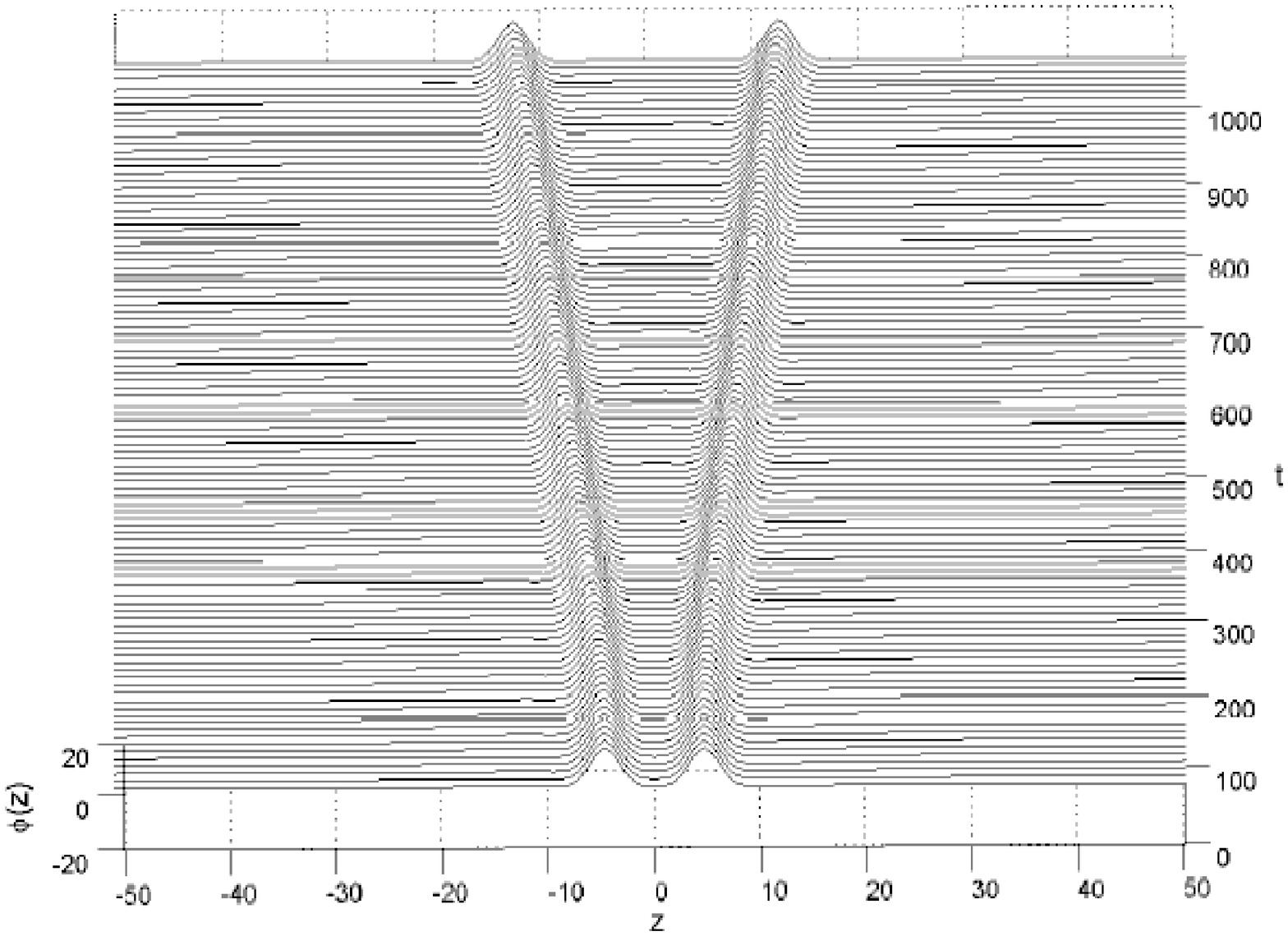}
\includegraphics[width = 7cm]{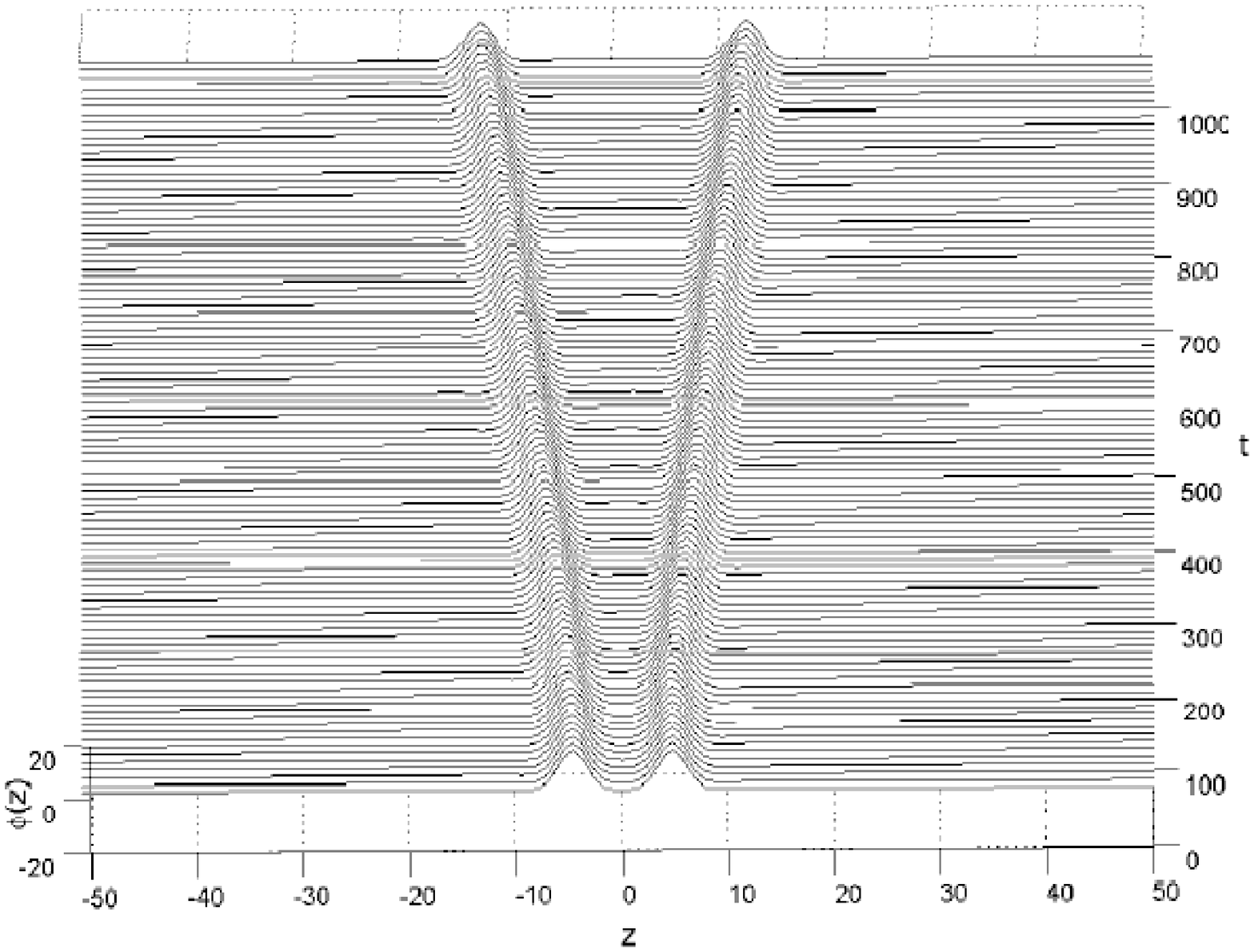}
\caption{Individual simulations of the initial-value problem
(\ref{kdv-Fourier}) and (\ref{initial-condition-PDE}) with $s = 2.3$
(top left), $s = 2.8$ (top right), $s = 3.6$ (middle left), $s =
4.2$ (middle right), $s = 4.5$ (bottom left) and $4.7$ (bottom
right).}
\end{center}
\end{figure}

\begin{figure}
\includegraphics [width = 14cm] {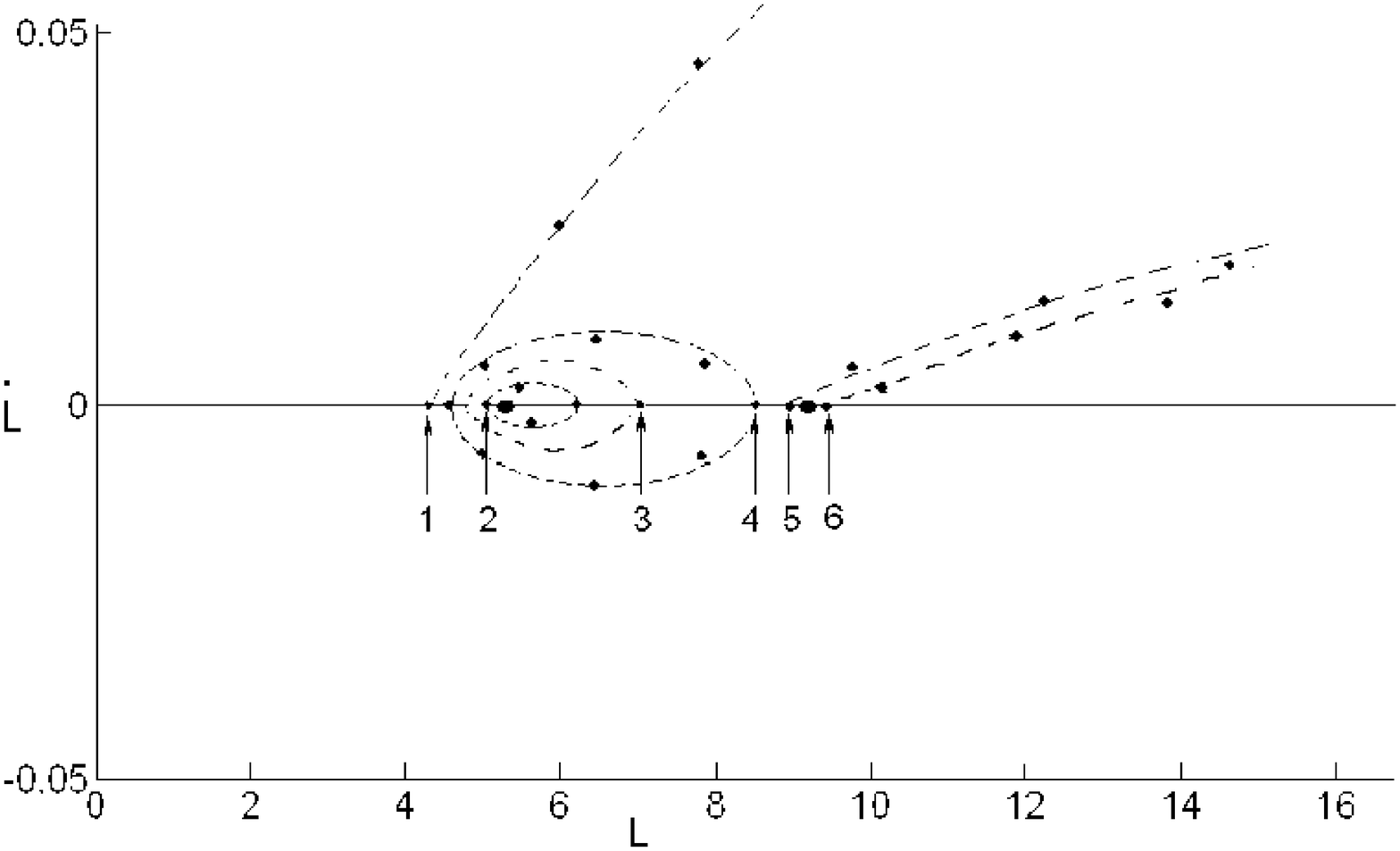}
\caption{The effective phase plane $(L,\dot{L})$ for simulations
on Figure 8, where $L$ is the distance between two pulses. The six
trajectories correspond to the six individual simulations in the
order described in the text. The black dots denote stable and
unstable equilibrium points which correspond to the two-pulse
solutions $\phi_1(x)$ and $\phi_2(x)$.}
\end{figure}

Summarizing, we have studied existence, spectral stability and
nonlinear dynamics of two-pulse solutions of the fifth-order KdV
equation. We have proved that the two-pulse solutions can be
numerically approximated by the Petviashili's iterative method
supplemented with a root finding algorithm. We have also proved
structural stability of embedded eigenvalues with negative Krein
signature and this result completes the proof of spectral
stability of two-pulse solutions related to the minima points of
the effective interaction potential. The validity of the Newton's
particle law is illustrated by the full numerical simulations of
the fifth-order KdV equation (\ref{kdv}) which show agreement of
slow nonlinear dynamics of two-pulse solutions with predictions of
the Newton's particle law.

{\bf Acknowledgement.} The authors thank Marcella Fioroni and Taras
Lakoba for contributions at the initial stage of the project.  The
research of M.C. is supported by the Sharcnet and NSERC graduate
scholarships. The research of D.P. is supported by the PREA and
NSERC grants.

\end{document}